# Ultra-Confinement of Polaritons in Single Atomic Layer Ag Photonic Quantum Dots

*Xinyi Li[1], Tetyana Ignatova[2], Chengye Dong[3,4], Krishnan Mekkanamkulam Ananthanarayanan[4], Rinu Abraham Maniyara[4], Arpit Jain[4], Furkan Turker[4], Vinay Kammarchedu[5], Aida Ebrahimi[4,5,6], Joshua A. Robinson[1,3,7,8], Slava V. Rotkin[1,8*]*

[1]Department of Engineering Science and Mechanics, The Pennsylvania State University, University Park, PA, USA
[2]Department of Nanoscience, Joint School of Nanoscience and Nanoengineering, the University of North Carolina at Greensboro, Greensboro, NC, USA
[3]Two-Dimensional Crystal Consortium, The Pennsylvania State University, University Park, PA, USA
[4]Department of Materials Science and Engineering, The Pennsylvania State University, University Park, PA, USA
[5]Department of Electrical Engineering, The Pennsylvania State University, University Park, PA, USA
[6]Department of Biomedical Engineering, The Pennsylvania State University, University Park, PA, USA
[7]Department of Chemistry, The Pennsylvania State University, University Park, PA, USA
[8]Department of Physics, The Pennsylvania State University, University Park, PA, USA

Light scattering by two-dimensional (2D) van der Waals heterostructures (vdWHs) is immense, especially given their infinitesimal volume, thus enabling strong light-matter interactions. Surface 2D polariton waves manifest through large concentration of electromagnetic field in vertical direction, normal to their propagation. By confining vdWH materials into 2D photonic shapes, one can manipulate and compress light in lateral directions. Scattering-type scanning near-field optical microscopy is a perfect tool for direct imaging of the propagating polaritons and studying the properties of confined polaritons in nanostructures. Though, thus far the quantitative analysis, such the wavelength extraction, has been challenged for confined polaritons by incapability of mapping of the wave period on sub-wavelength scale and difficulty of identifying an adequate substrate's "background" to subtract. Here, an analytical approach is developed to reveal the local propagation constant of confined polaritons under abovementioned constraints and map it with the sub-wavelength resolution. Applied to analysis of the SiC/2D-Ag/EG (epitaxial graphene) photonic nanostructures, the technique uncovered that the polaritons are highly confined in both vertical (~$\lambda$/50) and lateral directions (~$\lambda$/40) by 2D metal.

[*] e-mail: rotkin@psu.edu

## 1. Introduction

Light squeezing phenomenon in surface polariton and two-dimensional polariton (SP/2DP) waves, which manifests with the wavelength being much smaller than that of the bare light in vacuum, enables strong light-matter interactions. Since the photon energy density (and the light intensity) is inversely proportional to the wavelength, the light squeezing allows to achieve large fields, leads to non-linear physics and opens up possibilities for a number of fundamental studies in quantum optics as well as some novel applications. While the large energy density for propagating/free SPs/2DPs is due to the dielectric contrast at the interface, an additional light confinement can be reached through localization of the polaritons in plasmonic shapes, such as: photonic dots/metaatoms, metasurfaces, waveguides, nanoantennas and resonators, typically made of low-dissipation metals.[1–3] The range of materials used to fabricate plasmonic structures has been recently extended to two-dimensional materials (2DMs). Highly confined polaritons in nanostructures based on 2DMs and their van der Waals heterostructures (vdWHs) have gained increasing interest in recent years[4–17] with their strong light-matter interactions as well as a capability to manipulate the latter at the scales down to a single atomic layer.[18] Atomically thin 2DMs possess unique properties compared to their 3D counterparts[19] and, even without patterning, support various surface polaritons.[20–22] When integrated into vdWHs, they enable polariton confinement[20,23–25] which opens up exciting possibilities for developing next-generation optoelectronic,[23] photonic,[26,27] and sensing technologies.[28,29]

Here we demonstrate ultra-high confinement of mid-infrared (MIR) surface-phonon polaritons formed in the vicinity of a Reststrahlen band of polar SiC material. SiC samples are used as a substrate for confined heterostructure epitaxy (CHet) synthesis of 2D atomically thin layers of various metals and other materials, encapsulated by epitaxial graphene (EG), including but not limited to In, Ga, Sb.[30–33] In this work, a 2D-Ag monolayer is the material of choice due to its strong non-linear optical response.[34,35] These samples support an evanescent SP, confined along $z$, a normal direction, which is a composite SP mode of SiC/2D-Ag/EG vdWH. While it should propagate freely in the $x$-$y$ plane in the as-synthesized material, a lateral confinement is achieved by fabricating a photonic dot of the 2D-Ag/EG.

To evaluate plasmonic performance of the vdWH photonic dots, we use scattering Scanning Near-field Optical Microscopy (sSNOM)[20] to experimentally determine both the SP/2DP wavelength, $\lambda_{SP}$, and the confinement factor (CF). The CF is related to the leak of electromagnetic field outside of the photonic dot area which is defined by the magnitude of SP at two sides of the boundary of plasmonic nanostructure. The method developed here allows us to explicitly evaluate the spill-out of photonic wavefunction beyond the physical dimension of the confinement shape. We will show that this spill-out is negligible in SiC/2D-Ag/EG vdWH photonic dots of sub-micron size, being substantially smaller than the SP wavelength.

sSNOM is an ideal tool to map free SP waves and, thus, directly measure their dispersion relation,[20] that is, the wavevector $k_{SP}$ (and dissipation length as an imaginary component of the complex wavevector) as a function of SP frequency (sSNOM excitation frequency), as well as to detect nanoscale optical non-uniformities that could affect photonic properties.[29] However, a common approach[36] – hyperspectral mapping of the periodicity of standing wave fringes vs. the excitation wavelength, $\lambda_o$ – is limited to plasmonic systems larger than $\lambda_{SP}$. Since the size of the photonic dots in our samples is smaller than $\lambda_{SP}$, the sSNOM cannot visualize the full wave period[11] and the standard analysis fails to determine $k_{SP}$ and quantify the ***lateral*** light squeezing.

Furthermore, the wave periodic patterns in the sSNOM signal are combined with a background which has a negligible spatial dependence for uniform materials. However, this background varies substantially across the edge of the photonic structure, prohibiting the determination of actual SP wave amplitude difference from the sSNOM map and, thus, obscures definition of both the wave period and the CF. This is a notorious problem of proper referencing of the sSNOM signal, i.e., subtracting an unknown background signal.

To overcome this challenge, we developed an analytical approach for sSNOM hyperspectral mapping to reveal the local value of SP/2DP propagation constant and map it with deep-sub-wavelength resolution, only limited by the sSNOM noise level. This new method is based on eikonal[37] representation of the SP wave. By tracing the phase increment in the complex space of demodulated near-field signal and analyzing the trajectory of wave phasors (in Argand diagram of original sSNOM data), we determine the propagation constant of confined surface polaritons for 2D-Ag/EG plasmonic quantum dots with sub-wavelength size, lithographically fabricated on SiC substrate. The method will be demonstrated to simultaneously mitigate the referencing/normalization problem, by subtracting the spatially independent background and, thus, potentially enable optical nano-spectroscopy of polaritonic materials by taking a hyperspectral sequence for SP/2DP maps.

## 2. Results and Discussion

Cross-sectional scanning transmission electron microscopy (STEM) has been performed to determine morphology and material composition in the original SiC/2D-Ag/EG sample, before lithography. In **Figure 1**c, a typical cross-section of the vdWH sample shows that Ag forms a single atomic layer (of brightest contrast) between SiC substrate and EG, making up to 2 encapsulating graphene layers on the top. The photonic quantum dot structures were fabricated in vdWH material by e-beam lithography (see schematics of the process in Figure 1a) with a range of sizes, all smaller than 1 micrometer in diameter, as shown in the AFM topography image in Figure 1b. Black dashed lines indicate the boundaries of nano-disks as a guide for the eye. Surface morphology of the samples shows “snake-like” features[30] of ~0.7 nm height that are consistent with incomplete irregular second layer of EG.

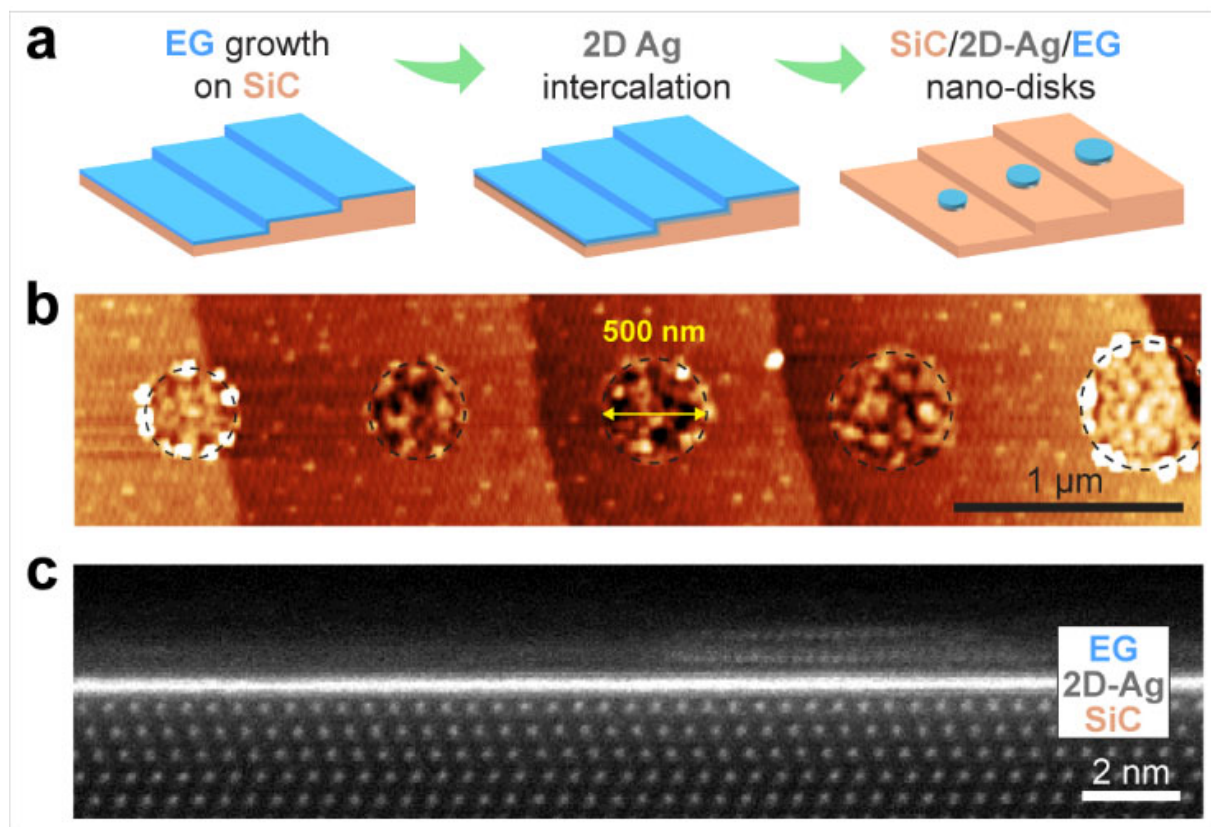

**Figure 1.** Material composition of 2D-Ag/EG plasmonic structures. (a) Illustration of the fabrication process for 2D-Ag/EG photonic quantum dots. (b) AFM topography of a series of disk-shape dots. Scale bar is 1 μm. (c) Representative STEM image of as-synthesized SiC/2D-Ag/EG material. Scale bar is 2 nm.

Hyperspectral sSNOM mapping of the region of interest – the central nano-disk of ~500 nm diameter, marked with a yellow arrow in the AFM image in Figure 1b and shown in a magnified AFM image in **Figure 2**b – was performed in the MIR range: 963-1040 $cm^{-1}$. Notably, this region covers the SiC LO phonon band. Figure 2 shows representative single-excitation-frequency images of the third-harmonic demodulated sSNOM raw signal: (c) optical amplitude, Abs($S_3$), and (d) optical phase, Arg($S_3$), compared to disk geometry from the AFM topography image in panel (b). While the fact of the SP confinement is seen by the brightness/contrast of the map – a higher (raw) optical amplitude signal inside the nano-disk compared to the outside region (substrate) – as well as via the phase contrast, the degree of this ***lateral*** confinement cannot be quantified from the Abs($S_3$) signal because it also contains an unknown background step at the edge, besides the SP wave amplitude contrast. We propose an approach to split the contributions from the SP wave and from the materials’ dielectric contrast by fitting the SP eikonal wave. Additionally, both the phase and amplitude maps allow one to trace a narrow belt around the disk edge where the signal is lower than in the disk center, though still much higher than that of the substrate (see Figure S1). This belt is likely a silver oxide material formed due to self-limited oxidation of the edge of 2D-Ag, which will be confirmed by SP eikonal analysis below. Such additional lateral heterostructures are often formed in 2D vdWH systems, which makes the standard SP analysis even harder. Markedly, there is no clear peak-to-trough pattern formed within the nano-disk due to its sub-diffraction size with respect to the SP wavelength, when the belt is excluded from the wave period analysis.

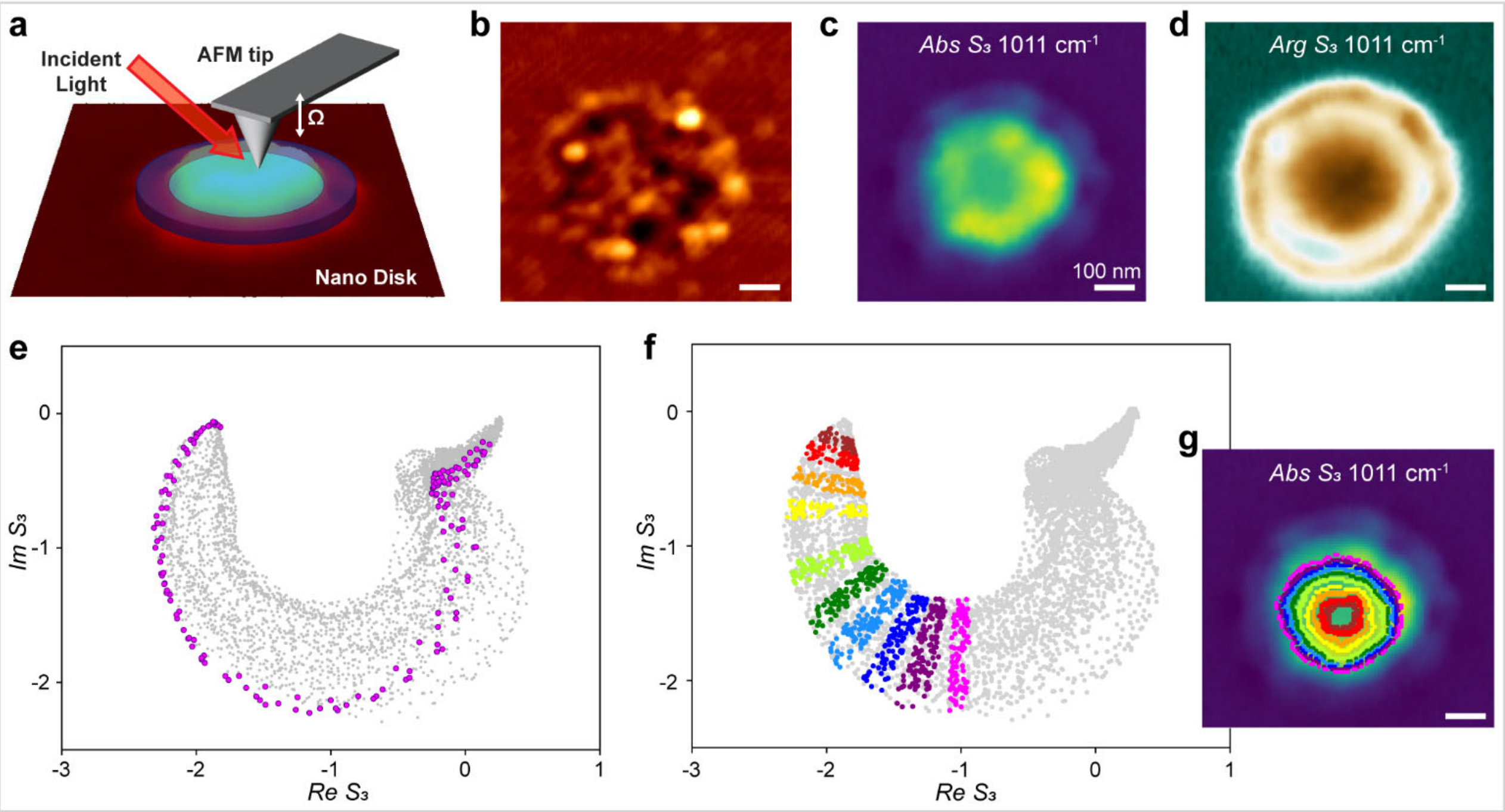

**Figure 2.** Surface polariton waves revealed by eikonal analysis. (a) Schematics of sSNOM imaging of the SiC/2D-Ag/EG nano-disks. (b) The AFM topography image; the raw maps of (c) third demodulated harmonic sSNOM optical amplitude, Abs($S_3$), and (d) optical phase, Arg($S_3$). (e-f) Argand plot: cross-correlation between Re($S_3$) and Im($S_3$) signals (purple dots in (e) are taken along fixed angular direction in real space image). (f-g) Correlation between the selected cluster points in Argand space with the same in spatial map of Abs($S_3$). Scale bars are 100 nm.

Figure 2e shows the Argand plot, i.e., the correlation plot between imaginary and real parts of the complex-valued $S_3$ sSNOM signal, taken from the whole map (grey points) at the excitation wavelength of 1011 cm$^{-1}$. In this Argand plot, a series of arc shapes is clearly seen. These arcs in the complex plane correspond to eikonal waves of a nearly constant amplitude and variable phase: $M\ e^{i\phi} = M\ e^{iks}$ where the phase evolution $\phi = k\ s$ is related to $k$, the SP wavevector and $s$, the distance along the propagation direction. For example, the trace of an evolution of the phasor (complex vector) of the $S_3$ signal along one of the radial directions from the center of plasmonic dot towards the substrate region is shown as a series of purple points.

Two adjacent arc segments of the phasor trajectory are seen, of a larger and small radius, that correspond to two SP eikonal waves in the 2D-Ag dot itself and in the oxide belt next to it, with different background (reference) signal, magnitude, velocity, and phase of SP modes (that correspond to arc center, radius, tangential derivative, and start phase, respectively). We emphasize that the true eikonal wave phase velocity computed along the phasor trajectory equals the local SP propagation constant *k*. Notably, this quantity can be computed, in ideal case of negligible noise, with the resolution of 2× pixel size of the raw sSNOM map, which is far beyond any other existing analytical near-field method.

To further prove that the phasor arcs in Argand plane correspond to different SP trajectories, we use the fact that in this particular map all arcs are nearly concentric (have a common center). Neglecting for a moment the difference in actual center position, i.e., temporarily postulating a constant background signal (a single center for all data points), we can define narrow regions along a fixed radial direction – “spokes” of different colors in Figure 2f. Such a set of points with nearly the same SP phase should correspond to the eikonal wavefront. Indeed, Figure 2g shows these points correspond to continuous circular curves in the real-space map, marked with the same color as the corresponding spoke. The radial shape for the confined SP eigenmode is typical for the axial symmetry of the nano-disk. We emphasize that the shape of the wavefronts, unlike the raw amplitude (or phase) signal, is robust to local defects and other perturbations due to the disk morphology, seen in the AFM image in panel (b). Additional evidence for negligible influence of the surface morphology on the sSNOM signal and, therefore, the eikonal analysis is provided in Supplementary Information (SI) section S7.

While the radial spoke of constant phase in Argand space corresponds to the wavefront in real space, the displacement along the arc trajectory in Argand space corresponds to the evolution of the eikonal phase, i.e., the wave propagation normal to the wavefront. By tracing the segments of an individual arc picked in the Argand space in SI Figure S3a, the SP wave evolution in radial direction is seen in Figure S3b. Note that the arc segments of approximately the same angular size (arc length) do not appear equally spaced in the map in panel (b). Indeed, this reflects on the change of the phase velocity along the radial direction: equidistant regions in real space in Figure S3d result in the variable incremental arc length of the segments of the same color in Figure S3c.

Based on these qualitative results, we fit all data points (of a given arc cluster) to a circle, as represented by the pink curve in **Figure 3**d. We stress that the center of the arc determines the natural reference for sSNOM calibration, bypassing the necessity to identify and subtract the unknown background signal. The numerical derivative of the phase along the arc with respect to the real-space coordinate along the eikonal wave propagation directly gives $k$, the propagation constant of the SP wave, with the spatial resolution of the order of twice of the maximum pixel size.

To be noted, the total arc spread across the photonic dot is about three quarters of π (less than π) in the map shown in Figure 3d. Thus, the overall distance that the SP wave propagates in real space is less than $\lambda_{SP}/2$, which explains why no complete fringe pattern (full period of SP wave) can be observed. For such a sub-diffractional size of the nano-disk, $\lambda_{SP}$ cannot be determined by any existing technique. Outstandingly, with the new method not only can we qualitatively reveal the existence of radial surface polaritons by the imaging of the near-field optical signal in the Argand space, but we can quantitatively measure the SP wavevector and, in principle, map it with the resolution limited by the sSNOM pixel size.

With the new technique, a “tail swirl”, a small segment in Argand plane due to the propagation of SP in the oxidized edge belt, which falls out of the single arc sequence, can be fitted with a separate SP eikonal wave (light green circle in Figure 3d and Figure S4a). Each eikonal SP has its own reference point (arc center), which means a different optical background for bare silver and for silver oxide materials to be subtracted. There is a large drop of the background across the oxide/bare Ag boundary, even though the sSNOM amplitude is continuous (arc trajectory is not broken). Without new eikonal wave analysis, the sSNOM amplitude or phase, Abs($S_3$) or Arg($S_3$), cannot be used to determine the edge of two distinct materials. The physical origin of the background sSNOM signal is due to the surface impedance of the displacement current flowing from the tip to the sample, neglecting (or averaging out) the SP propagation and interference patterns. Here, a different value of surface impedance of the SiC/2D-Ag/EG vdWH for bare Ag-dot and for oxidized silver region is detected, allowing one to clearly trace the oxide layer boundary (Figure 3b-c).

A smaller radius of the oxide SP’s arc corresponds to a substantially smaller magnitude of the SP wave, which is the quantitative measure of the lateral field confinement. Upon closer examination, we also noticed that the SP wave exists outside of the photonic dot region with much smaller, though non-zero magnitude, shown as a short blue arc in the zoomed-in Argand plot in Figure S4b. This radius of the arc in the SiC substrate, ~0.002, is far smaller than the SP magnitude in the oxide, 0.396, which is smaller than that in bare 2D-Ag material, 1.471, at the excitation wavelength of 994 $cm^{-1}$. Thus, the total ratio of the sSNOM field between the photonic dot central region and the substrate outside the oxide belt constitutes >700.

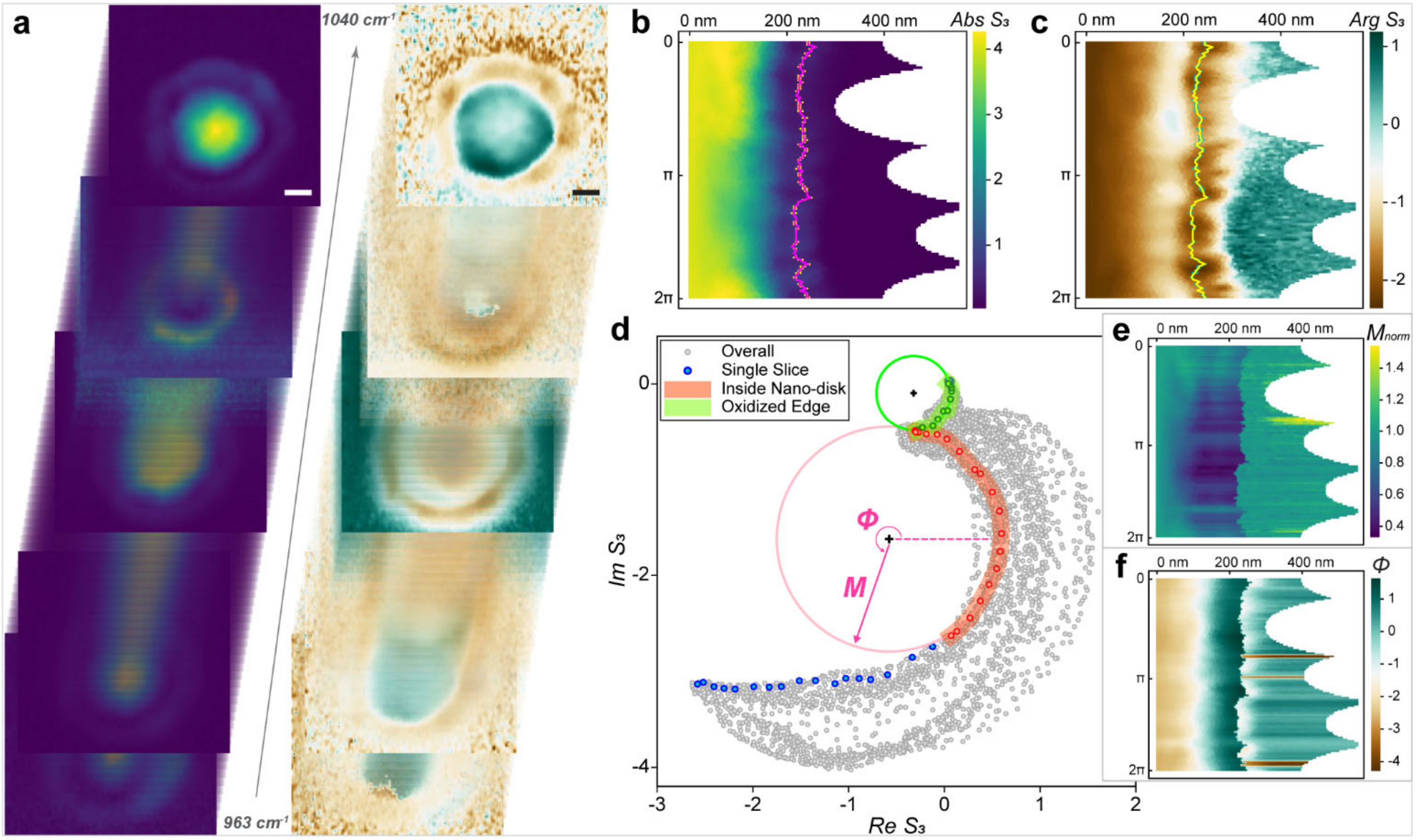

**Figure 3.** Spectral analysis of eikonal wave model. (a) Sketch of the hyperspectral sSNOM mapping using excitation wavelength 963-1040 cm$^{-1}$. Scale bars are 100 nm. (b-c) Representative sSNOM Abs($S_3$) and Arg($S_3$) maps at excitation wavelength 994 cm$^{-1}$, after polar coordinate transformation. (d) Argand space representation of data from (b-c); the datapoints selected along a single radial direction are fitted to two eikonal curves (red and green scatter dots and arcs, the centers are marked by black crosses). The coordinates of adjacent points between two eikonal waves are shown in panels (b-c) as magenta/yellow dots. (e-f) Fitted maps of true magnitude ($M$) and referenced phase angle ($\phi$) for eikonal waves.

The propagation constant derived with our eikonal phase analysis allows to determine the SP dispersion even in the absence of a clear wave pattern formed, e.g., for confined modes. The flowchart of the analysis is shown schematically in Figure 3, and detailed in Methods section. In brief, each pair of a single frequency maps (Abs and Arg, panels (b) and (c)) is fitted to a radial eikonal wave and a background component, as shown in panel (d) for a single radial direction (evolution of SP along the radial coordinate from the center of nano-disk). The fitted parameters for all radial vectors are combined into the new maps for true SP eikonal magnitude and phase (panels (e-f)). Note the sharp boundary in eikonal parameters at ~250 nm radial coordinate, which corresponds to the inner edge of oxide belt, as opposed to the broad features in original data, displaced from the actual belt edge location.

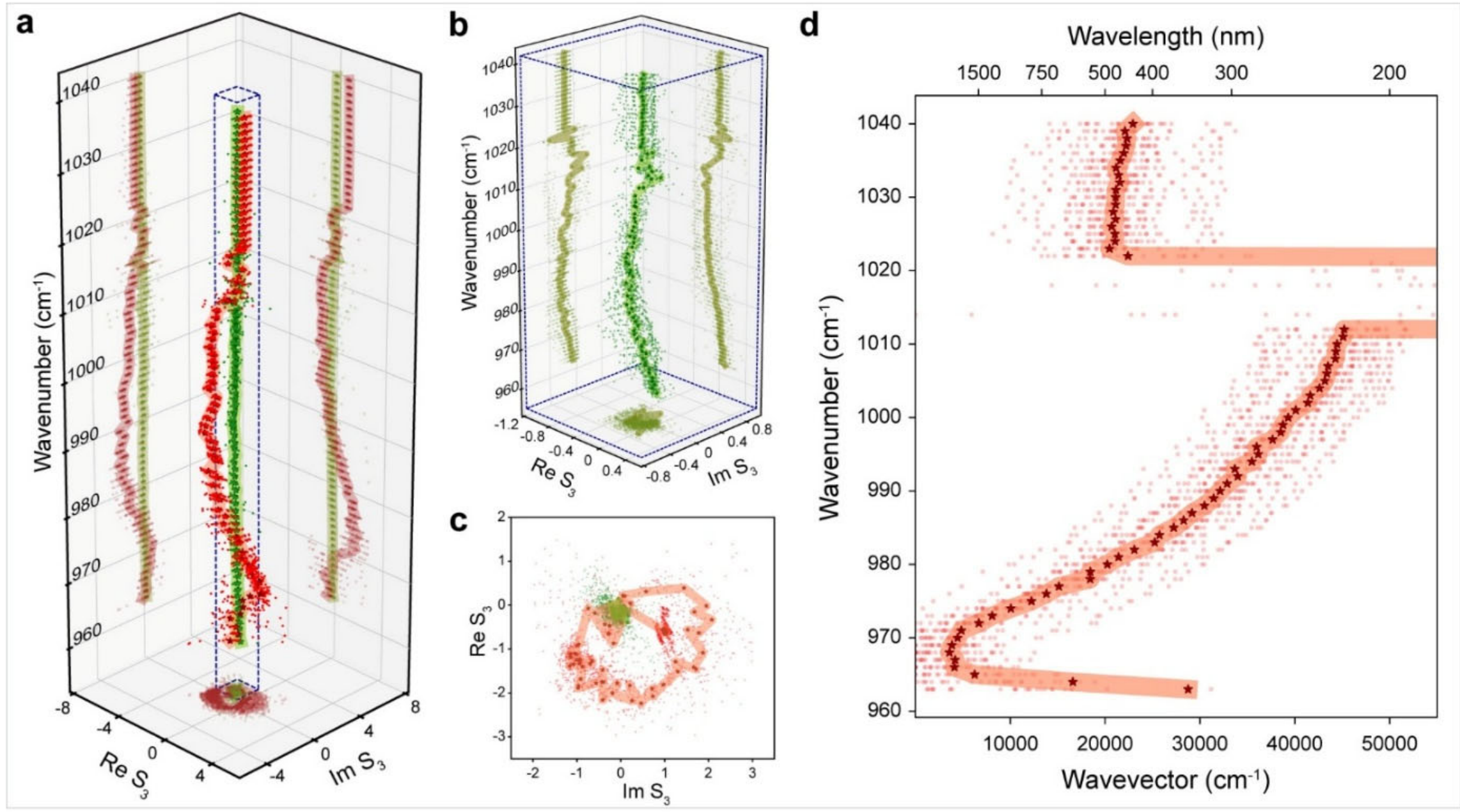

**Figure 4.** Dispersion of material properties of bare Ag vs. oxidized belt. (a) 3D hyperspectral Argand plot of background (center coordinates) of the fitted bare Ag (red) and oxide (green) cluster arcs from Figure 3d vs. excitation frequency (side walls show projections on Real and Imaginary axes). (b) Zoomed-in view of the oxide belt data (green) from the box in panel (a). (c) 2D Argand plot of the data in panel (a). (d) Dispersion relation of the SP as computed along different radial directions (the line shows the average dispersion relation for the whole dot).

The background signal (sSNOM reference) varies non-monotonously with the excitation frequency, as shown in **Figure 4**a. Figure 4c presents the Argand plot of the frequency evolution of the reference (not to be mixed with the spatial evolution of SP phasor in Argand plots in the rest of the paper). One should expect non-negligible frequency dependence in the vicinity of LO mode of the SiC substrate. Although, a complicated response of the vdWH with multiple 2D material components needs to be modeled to predict an explicit shape of optical impedance, some qualitative arguments are presented next. We stress that the spectral dependence of the reference is due to the non-propagating (non-polariton) near-field components, the main part of which is the uniform dielectric screening by the sample. If one considers this referencing to be due to the screening of the tip (image dipole) by the vdWH stack, a full 2π cycle of the reference in complex plane corresponds to passing a pole of the dielectric response function. We observed that the reference (i.e., the central point of the fitted arcs in Figure 3d) for oxide (green cluster) varies much less with the frequency compared to the reference point for bare Ag (red cluster arc), though both have similar spectral features, which becomes clearer in the zoomed-in image in Figure 4b. While the spectral behavior in the excitation range of 963-1040 cm$^{-1}$ is formed by the optical phonons of SiC, as discussed next, the optical responses of the regions of distinct 2D material – the bare Ag inside nano-disk and the oxidized belt – heavily modulate the signal. Being less conductive than metallic silver, the SiC/oxide/EG structure shows less screening, which is consistent with a smaller confined SP amplitude.

The propagation constant for SP was obtained from the eikonal fitting of the hyperspectral cube of sSNOM maps vs. the excitation frequency from 963 to 1040 cm$^{-1}$. The resulting SP dispersion relation for bare Ag is shown in Figure 4d, averaged over different radial directions. The main feature (discontinuity) of the dispersion plot is due to the Reststrahlen band of SiC optical phonon (see also hyperspectral data on "bulk" non-patterned SiC/2D-Ag/EG films in SI, Section S9). Notably, similar features appeared in the background signal in panels (a, b) in the same spectral region.

It is instructive to compare results of the eikonal wave model to the "classical" approach (by counting peaks in sSNOM map to define a wavelength). Although, this cannot be done in sub-diffractional pQDs of our study. In order to validate the new eikonal technique, in Section S8 of SI, we provide a similar study on hBN flakes with the area much exceeding the polariton wavelength. Figure S14 proivides the summary of results: the eikonal technique allows to experimentally obtain the polariton dispersion curve in better agreement with theoretically predicted one, also using data from a local area, significantly smaller than needed for peak counting (cf. number of pixels used, ~40 in eikonal model vs. ~200, as shown in Figure S12b,c).

Further validation of the dispersion of the confined polariton can be obtained from a comparison to the bulk SP modes. In section S9 of the SI, we present the hyperspectral mapping of non-structured SiC/2D-Ag/EG films, that is the sample which was not used for nanofabrication of plasmonic quantum dots. A common wavelength analysis was not possible for detection of the SP waves due to very strong diffraction of the (long wavelength) modes off the surface morphological non-uniformities. As we discuss in the Introduction, standard tools are often not capable to determine the SP wavelength which motivated us to develop an alternative spectroscopic analytical technique here. The Fourier analysis was implemented to receive some spectral information about bulk SP modes, without spatial resolution (averaged over the area of several sq. micrometers), as detaild in SI. Even though not concluding, the spectral results on bulk samples can be compared to the dispersion in Figure 4d. In brief, confined and free (bulk) modes share a number of common spectral features (see Figure S17 and S19 and Section S9), due to the same physics of MIR response of SiC/2D-Ag/EG material.

However, it is the eikonal model which allows to get the SP dispersion even for the samples with the complex structure.

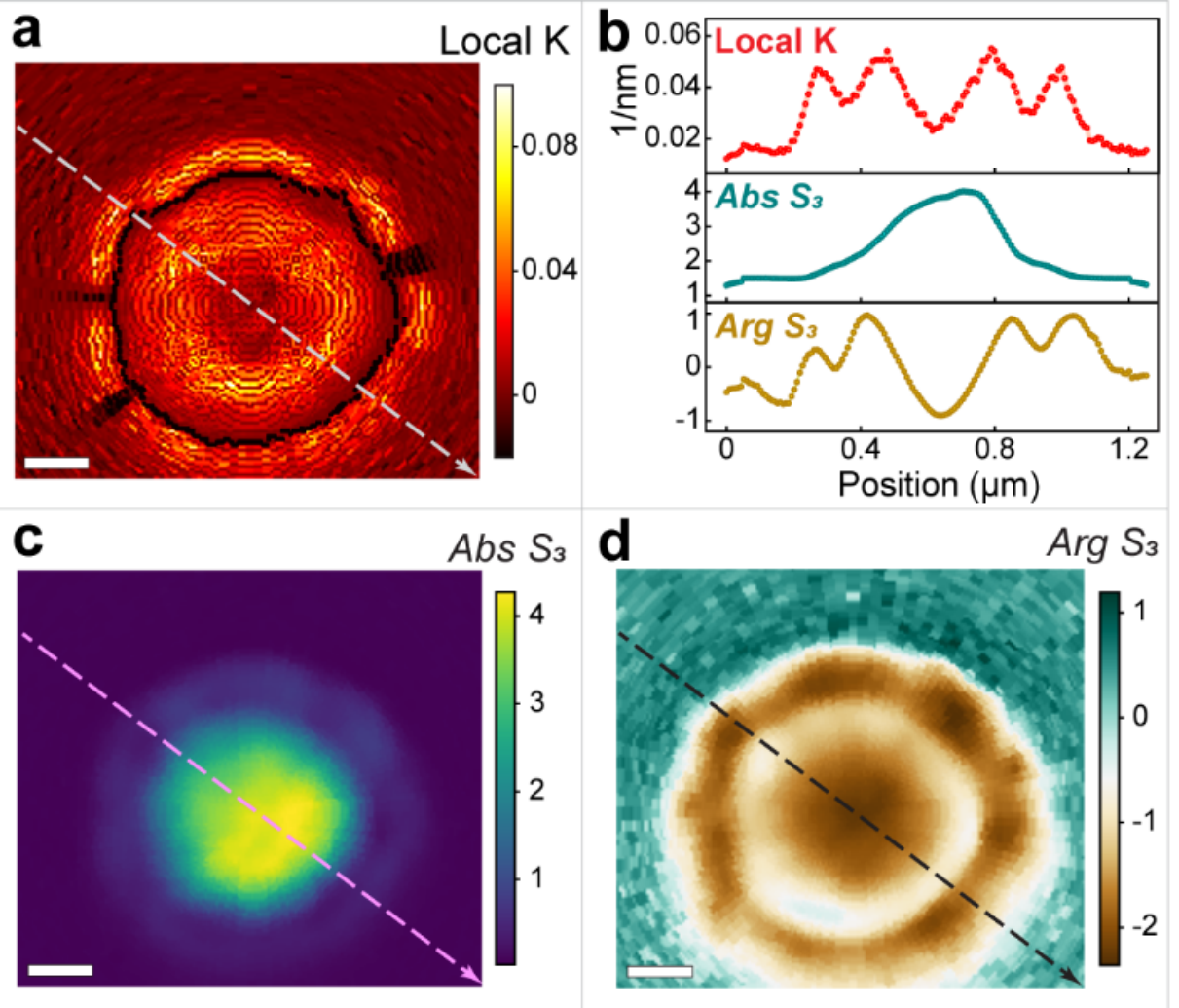

**Figure 5.** Local propagation constant for confined SP. (a) Calculated map of local propagation constant. (b) Line profiles of the local propagation constant, Abs($S_3$) and Arg($S_3$) (from top to bottom) extracted from the diagonal line in each map. (c-d) sSNOM Abs($S_3$) and Arg($S_3$) maps for comparison. The excitation wavelength equals 994 $cm^{-1}$. Scale bars are 100 nm.

Within the eikonal model, the map of $\frac{\partial \phi}{\partial r}$ shows the local values of the propagation constant of the SP wave (**Figure 5**a). To verify the applicability of the radial SP wave model, we also derived the other components of the gradient of the eikonal wave (presented in Figure S6a-d in polar, not Cartesian coordinates). Note that all other components of the gradient are much less, at the noise level, compared to the radial propagation constant. The eikonal technique allows for local mapping of the wavevector, independent of the background, as one can see from the line profiles of $k_{SP}$, Abs($S_3$), and Arg($S_3$) in Figure 5b, extracted from the line shown in Figure 5a, c, and d. The non-flat shape of the *k*-profile is expected for a confined mode, being a wave packet, as opposed to the free SP, with a single value of its wavevector for fixed SP frequency.

In this work, a very strong **space localization** ratio was measured for SiC SPs confined by 2D-Ag/EG: indeed, the photonics quantum dot radius is ~250 nm < $\lambda_o/40$, for $\lambda_o$~10 μm. The complete SP localization with no traceable field spill-out outside of the deep-sub-wavelength dot shape was achieved: we estimate the true SP wave magnitude ratio >150× to appear between the photonic dot edge (to remember this is natural Ag-oxide belt area) with respect to the bare SiC substrate outside the confinement area, while an additional factor comes from the ratio between the dot center region and the oxide belt ranging from 3 to 16×, maximized at 973 cm$^{-1}$, around the LO feature of SiC near 966 cm$^{-1}$. This results in an extreme field lateral confinement factor ~40, along with the large vertical CF=$\lambda_o/\lambda_{SP}$~50. Large CF factors were obtained earlier by detecting the SP wavelength at the edge of other 2D materials: the hBN, $Ge_3Sb_2Te_6$, and transition metal dichalcogenide (TMDC) samples, as summarized in **Table 1**.

**Table 1.** Confinement factors reported in other 2DMs or 2D vdWHs with MIR excitation.

| Material | $\lambda_o/\lambda_{SP}$ | Lateral CF | Refs |
|---|---|---|---|
| Few-layer $MoO_3$ | 120 | -- | [21] |
| hBN/Au grating | 132 | -- | [22] |
| Monolayer graphene (MLG) in hBN | 150 | -- | [24] |
| $MoS_2$, $MoSe_2$, $WS_2$, $WSe_2$/SiC | up to 190 | -- | [25] |
| hBN cones | 86 | 61 | [5] |
| $MoO_3$ nano-ribbon | 133 | ~60 | [15] |
| MLG/hBN nanodisks | ~100 | ~48 | [16] |
| hBN/Au cavity | ~50 | ~24 | [17] |

## 3. Conclusion

In conclusion, an analytical method for sSNOM hyperspectral mapping was developed to reveal the local values of the propagation constant of confined surface phonon polaritons. The method was demonstrated on SP localized in 2D-Ag/EG plasmonic nano-disks with sub-wavelength sizes. Based on the new eikonal wave approximation, we detected an ultra-strong confinement of SiC polariton waves by 2D-vdWH-based photonic quantum dots. The spectral dependence of both the phase velocity (wavevector) and background sSNOM signal (dielectric screening response function) reveals the Reststrahlen band of SiC as the physical origin of the effect. Scanning the trajectory of SP phasors (in Argand diagram of the sSNOM data), we confidently defined the boundary between regions with different material composition of the nano-disk, the bare 2D-Ag center and a belt due to partial oxidation of the silver edge, which cannot be easily revealed by other optical characterization techniques. Finally, the map of local SP propagation constant was obtained, with substantially sub-wavelength resolution, which allowed us to measure the lateral confinement of vdWH SPs. Extreme values of CF, which is due to just a few layers of a composite 2D material, witness the large photon density of states near the surface, thus opening opportunities for engineering non-linear and quantum 2D-photonic devices in the future.

## 4. Methods

***Sample Fabrication.*** The fabrication process is schematically illustrated in Figure 1a. Monolayer epitaxial graphene (EG) was first grown on a 6H-SiC(0001) substrate (Coherent Corp.) via thermal decomposition. The SiC surface was annealed at 1500 ℃, 700 Torr in a 10% $H_2$/Ar mixture for 30 min, followed by heating to 1800 °C in pure Ar (700 Torr, 20 min) to form monolayer EG.

Subsequently, a monolayer of Ag was prepared through confinement heteroepitaxy (CHet). The EG surface was treated by $O_2$ plasma to generate defects that act as intercalation pathways. The plasma-treated EG was then placed face-down in a crucible containing ~50 mg of Ag powder, followed by annealing at 900 °C, 500 Torr, with 50 sccm Ar carrier gas for 1 h, enabling Ag intercalation at the EG/SiC interface.

Circular discs of 2D-Ag/EG were defined by electron-beam lithography (EBL). The substrate was spin-coated with PMMA and baked at 180 °C for 90 s, exposed using a Raith EBPG5200 Plus e-beam system, and developed in a 1:1 mixture of MIBK:IPA for 60 s. The exposed 2D-Ag/EG regions were etched by $N_2$ plasma (20 °C, 15 s) in an ULVAC NE-550 system. Finally, the resist was removed by lift-off in PRS3000, acetone, and IPA, yielding isolated 2D-Ag/EG nanodisc arrays.

We assume that the formed 2D Ag to be mostly one-layer based on the predictions carried out in previous study where the relationship between the number of layers and metal's chemical potential was established based on the first-principles equilibrium-phase stability calculations.[30] Advanced calculation predicts formation of two distinct phases with slightly different Ag density, depending on the registration with underlying SiC lattice.[38] The SiC morphology should also determine the EG layer thickness (up to 2 layers) and continuity. The areas outside the plasmonic nanostructures, after etching, are expected to be bare SiC, as illustrated in the last step of fabrication process schematic picture.

***Sample Characterization.*** Cross-sectional specimens were prepared using in-situ lift-out in a focused ion beam (FIB) system (Helios Nanofab DualBeam 660, FEI Inc.). To minimize damage and contamination during milling, an initial ~20 nm amorphous carbon layer was deposited on the sample surface by sputter coating. Subsequently, an additional ~400 nm amorphous carbon protection layer was deposited sequentially by electron-beam deposition followed by ion-beam deposition prior to milling. FIB thinning was performed with a $Ga^+$ ion beam, starting at 30 kV and gradually reduced to 1 kV to achieve fine thinning of the lamella. High-resolution scanning transmission electron microscopy (STEM) was performed using an FEI Titan³ G2 microscope operated at an accelerating voltage of 200 kV. The probe was configured with a convergence semi-angle of ~30 mrad and a probe current of ~70 pA. High-angle annular dark-field (HAADF) images were acquired using a detector with a collection angle range of 51-300 mrad, providing Z-contrast sensitivity to the atomic structure.

The near-field imaging was performed using a scattering-type scanning near-field optical microscope, a custom-built Neaspec system. The instrument operated in pseudo-heterodyne mode with a tapping amplitude of ~70 nm, using ARROW-NCPt probes from Nanoworld (tip radius < 25 nm). For hyperspectral mapping, continuous-wave quantum cascade laser excitation (MIRCat, Daylight Solutions) was used in the range of 963-1040 $cm^{-1}$ (10.384-9.615 μm), with power kept below 2 mW at the focal aperture.

***Phase Analysis of SP Dispersion.*** A hyperspectral cube of sSNOM maps vs. the excitation frequency from 963 to 1040 $cm^{-1}$ (see SI Table 1) was carefully registered for batch processing. For each excitation wavelength, we obtain the reference point/background signal and the propagation constant for each of 3 separate regions: bare 2D-Ag central area, oxide layer belt, and SiC substrate (highlighted in red, green, and light blue in Figure S4). For the axial symmetry of the photonic dot, which yields the radial eikonal wave, we transform the real-space image from the Cartesian into polar coordinates. We generate the phasor map in Argand space. Within the nano-disk, data points with the same $r$ but different $\theta$ (polar coordinates) have approximately the same phase (belong to the single circular wavefront), compare to Figure 2f. The data points with the same $\theta$ must show fast phasor rotation with $r$, as highlighted in Argand plot (Figure 2e) by purple scatter points, corresponding to green rectangular region in the real space map, see Figure S2a. Radial direction of the maximum variation of the phase is used to compute SP propagation constant, noting a slow variation of the amplitude due to edge-induced perturbation.

The fit for the radial direction slices (both red and green data points) is performed to confine data to a circle in the Argand space. The turning point between two arcs, or preferably, the point where the change of phase flips the sign (with respect to the center of the red arc) defines the exact position of the boundary between pure 2D-Ag and oxidized edge belt (shown as magenta curve in Figure 3b). The center of fitted circle is the referencing point for each region. Note that the references of different regions vary for different excitation wavelengths.

Having a reference point allows to determine the true phase (and magnitude) of the phasor of the surface wave and overcome the ambiguity of proper referencing/normalization of sSNOM signals. This yields mapping of $M$ (Figure 3e), and $\phi$ (Figure 3f), the eikonal wave magnitude and phase, to the real-space polar coordinates $\theta$ and $r$, and stack it vs. the excitation wavelength for studying the SP dispersion. The gradient of the phase matrix with respect to the $\theta$ and $r$ will provide the propagation constant in chosen direction at each location in real space, at the fixed excitation wavelength. For the radial eikonal wave, the radial component of the gradient gives the wavevector of polariton, while the other component of gradient is negligible.

Similar eikonal analysis applied to sSNOM data collected from a series of photonic quantum dots of variable radius showed a close correlation for the measured propagation constants, as shown in Figure S8.

**Acknowledgements**

X.L., C.D., K.A, R.M., A.J., J.A.R. and S.V.R. acknowledge financial support from the Penn State MRSEC Center for Nanoscale Science via NSF award DMR2011839. C.D., J.A.R. and S.V.R. are also supported by 2DCC-MIP under NSF cooperative agreement DMR-1539916 and DMR-2039351. F.T. acknowledge the funding from the Air Force Office of Scientific Research (AFOSR) through contract FA9550-19-1-0295. TI acknowledge the Joint School of Nanoscience and Nanoengineering, a member of the South-Eastern Nanotechnology Infrastructure Corridor (SENIC) and National Nanotechnology Coordinated Infrastructure (NNCI), supported by the NSF (Grant ECCS-2025462).

**References**

[1] J. J. Baumberg, J. Aizpurua, M. H. Mikkelsen, D. R. Smith, *Nat Mater* 2019, *18*, 668.
[2] R. A. Maniyara, D. Rodrigo, R. Yu, J. Canet-Ferrer, D. S. Ghosh, R. Yongsunthon, D. E. Baker, A. Rezikyan, F. J. García de Abajo, V. Pruneri, *Nat Photonics* 2019, *13*, 328.
[3] T. Neuman, P. Alonso-González, A. Garcia-Etxarri, M. Schnell, R. Hillenbrand, J. Aizpurua, *Laser Photon Rev* 2015, *9*, 637.
[4] H. Yan, T. Low, W. Zhu, Y. Wu, M. Freitag, X. Li, F. Guinea, P. Avouris, F. Xia, *Nat Photonics* 2013, *7*, 394.
[5] J. D. Caldwell, A. V. Kretinin, Y. Chen, V. Giannini, M. M. Fogler, Y. Francescato, C. T. Ellis, J. G. Tischler, C. R. Woods, A. J. Giles, M. Hong, K. Watanabe, T. Taniguchi, S. A. Maier, K. S. Novoselov, *Nat Commun* 2014, *5*, 5221.
[6] Z. Fei, M. D. Goldflam, J.-S. Wu, S. Dai, M. Wagner, A. S. McLeod, M. K. Liu, K. W. Post, S. Zhu, G. C. A. M. Janssen, M. M. Fogler, D. N. Basov, *Nano Lett* 2015, *15*, 8271.
[7] A. Y. Nikitin, P. Alonso-González, S. Vélez, S. Mastel, A. Centeno, A. Pesquera, A. Zurutuza, F. Casanova, L. E. Hueso, F. H. L. Koppens, R. Hillenbrand, *Nat Photonics* 2016, *10*, 239.

[8] Z.-B. Zheng, J.-T. Li, T. Ma, H.-L. Fang, W.-C. Ren, J. Chen, J.-C. She, Y. Zhang, F. Liu, H.-J. Chen, S.-Z. Deng, N.-S. Xu, *Light Sci Appl* 2017, *6*, e17057.
[9] F. Hu, Y. Luan, Z. Fei, I. Z. Palubski, M. D. Goldflam, S. Dai, J.-S. Wu, K. W. Post, G. C. A. M. Janssen, M. M. Fogler, D. N. Basov, *Nano Lett* 2017, *17*, 5423.
[10] I.-H. Lee, D. Yoo, P. Avouris, T. Low, S.-H. Oh, *Nat Nanotechnol* 2019, *14*, 313.
[11] W. S. Hart, V. Panchal, C. Melios, W. Strupiński, O. Kazakova, C. C. Phillips, *2d Mater* 2019, *6*, 021003.
[12] S.-J. Yu, Y. Jiang, J. A. Roberts, M. A. Huber, H. Yao, X. Shi, H. A. Bechtel, S. N. Gilbert Corder, T. F. Heinz, X. Zheng, J. A. Fan, *ACS Nano* 2022, *16*, 3027.
[13] B. S. Y. Kim, A. J. Sternbach, M. S. Choi, Z. Sun, F. L. Ruta, Y. Shao, A. S. McLeod, L. Xiong, Y. Dong, T. S. Chung, A. Rajendran, S. Liu, A. Nipane, S. H. Chae, A. Zangiabadi, X. Xu, A. J. Millis, P. J. Schuck, Cory. R. Dean, J. C. Hone, D. N. Basov, *Nat Mater* 2023, *22*, 838.
[14] H. Ling, A. Manna, J. Shen, H.-T. Tung, D. Sharp, J. Fröch, S. Dai, A. Majumdar, A. R. Davoyan, *Optica* 2023, *10*, 1345.
[15] Y. Zeng, T. Sun, R. Chen, W. Ma, Q. Yan, D. Lu, T. Qin, C. Hu, X. Yang, P. Li, *Opt Express* 2023, *31*, 28010.
[16] Y. Luo, J.-H. Park, J. Zhu, M. Tamagnone, F. Capasso, T. Palacios, J. Kong, W. L. Wilson, *ACS Nano* 2024, *18*, 17492.
[17] H. Herzig Sheinfux, L. Orsini, M. Jung, I. Torre, M. Ceccanti, S. Marconi, R. Maniyara, D. Barcons Ruiz, A. Hötger, R. Bertini, S. Castilla, N. C. H. Hesp, E. Janzen, A. Holleitner, V. Pruneri, J. H. Edgar, G. Shvets, F. H. L. Koppens, *Nat Mater* 2024, *23*, 499.
[18] Y. Wu, J. Duan, W. Ma, Q. Ou, P. Li, P. Alonso-González, J. D. Caldwell, Q. Bao, *Nature Reviews Physics* 2022, *4*, 578.
[19] Y.-C. Lin, R. Torsi, R. Younas, C. L. Hinkle, A. F. Rigosi, H. M. Hill, K. Zhang, S. Huang, C. E. Shuck, C. Chen, Y.-H. Lin, D. Maldonado-Lopez, J. L. Mendoza-Cortes, J. Ferrier, S. Kar, N. Nayir, S. Rajabpour, A. C. T. van Duin, X. Liu, D. Jariwala, J. Jiang, J. Shi, W. Mortelmans, R. Jaramillo, J. M. J. Lopes, R. Engel-Herbert, A. Trofe, T. Ignatova, S. H. Lee, Z. Mao, L. Damian, Y. Wang, M. A. Steves, K. L. Knappenberger, Z. Wang, S. Law, G. Bepete, D. Zhou, J.-X. Lin, M. S. Scheurer, J. Li, P. Wang, G. Yu, S. Wu, D. Akinwande, J. M. Redwing, M. Terrones, J. A. Robinson, *ACS Nano* 2023, *17*, 9694.
[20] D. N. Basov, M. M. Fogler, F. J. García De Abajo, *Science (1979)* 2016, *354*.
[21] Z. Zheng, J. Chen, Y. Wang, X. Wang, X. Chen, P. Liu, J. Xu, W. Xie, H. Chen, S. Deng, N. Xu, *Advanced Materials* 2018, *30*.
[22] I.-H. Lee, M. He, X. Zhang, Y. Luo, S. Liu, J. H. Edgar, K. Wang, P. Avouris, T. Low, J. D. Caldwell, S.-H. Oh, *Nat Commun* 2020, *11*, 3649.
[23] X. Guo, W. Lyu, T. Chen, Y. Luo, C. Wu, B. Yang, Z. Sun, F. J. G. de Abajo, X. Yang, Q. Dai, *Advanced Materials* 2023, *35*.
[24] A. Woessner, M. B. Lundeberg, Y. Gao, A. Principi, P. Alonso-González, M. Carrega, K. Watanabe, T. Taniguchi, G. Vignale, M. Polini, J. Hone, R. Hillenbrand, F. H. L. Koppens, *Nat Mater* 2015, *14*, 421.
[25] A. M. Dubrovkin, B. Qiang, H. N. S. Krishnamoorthy, N. I. Zheludev, Q. J. Wang, *Nat Commun* 2018, *9*, 1762.
[26] F. Xia, H. Wang, D. Xiao, M. Dubey, A. Ramasubramaniam, *Nat Photonics* 2014, *8*, 899.
[27] M. Turunen, M. Brotons-Gisbert, Y. Dai, Y. Wang, E. Scerri, C. Bonato, K. D. Jöns, Z. Sun, B. D. Gerardot, *Nature Reviews Physics* 2022, *4*, 219.
[28] S.-H. Oh, H. Altug, X. Jin, T. Low, S. J. Koester, A. P. Ivanov, J. B. Edel, P. Avouris, M. S. Strano, *Nat Commun* 2021, *12*, 3824.

[29] T. Ignatova, S. Pourianejad, X. Li, K. Schmidt, F. Aryeetey, S. Aravamudhan, S. V. Rotkin, *ACS Nano* 2022, *16*, 2598.
[30] N. Briggs, B. Bersch, Y. Wang, J. Jiang, R. J. Koch, N. Nayir, K. Wang, M. Kolmer, W. Ko, A. De La Fuente Duran, S. Subramanian, C. Dong, J. Shallenberger, M. Fu, Q. Zou, Y.-W. Chuang, Z. Gai, A.-P. Li, A. Bostwick, C. Jozwiak, C.-Z. Chang, E. Rotenberg, J. Zhu, A. C. T. van Duin, V. Crespi, J. A. Robinson, *Nat Mater* 2020, *19*, 637.
[31] W. Lee, Y. Wang, W. Qin, H. Kim, M. Liu, T. N. Nunley, B. Fang, R. Maniyara, C. Dong, J. A. Robinson, V. H. Crespi, X. Li, A. H. MacDonald, C.-K. Shih, *Nano Lett* 2022, *22*, 7841.
[32] A. Vera, B. Zheng, W. Yanez, K. Yang, S. Y. Kim, X. Wang, J. C. Kotsakidis, H. El-Sherif, G. Krishnan, R. J. Koch, T. A. Bowen, C. Dong, Y. Wang, M. Wetherington, E. Rotenberg, N. Bassim, A. L. Friedman, R. M. Wallace, C. Liu, N. Samarth, V. H. Crespi, J. A. Robinson, *ACS Nano* 2024, *18*, 21985.
[33] C. Dong, L.-S. Lu, Y.-C. Lin, J. A. Robinson, *ACS Nanoscience Au* 2024, *4*, 115.
[34] C. Ordonez, C. Dong, A. Jain, L.-S. Lu, J. A. Robinson, K. L. Knappenberger, *The Journal of Physical Chemistry C* 2025, *129*, 5133.
[35] K. Zhang, R. A. Maniyara, Y. Wang, A. Jain, M. T. Wetherington, T. T. Mai, C. Dong, T. Bowen, K. Wang, S. V. Rotkin, A. R. Hight Walker, V. H. Crespi, J. Robinson, S. Huang, *Sci Adv* 2025, *11*.
[36] S. Dai, Z. Fei, Q. Ma, A. S. Rodin, M. Wagner, A. S. McLeod, M. K. Liu, W. Gannett, W. Regan, K. Watanabe, T. Taniguchi, M. Thiemens, G. Dominguez, A. H. C. Neto, A. Zettl, F. Keilmann, P. Jarillo-Herrero, M. M. Fogler, D. N. Basov, *Science (1979)* 2014, *343*, 1125.
[37] M. Born, E. Wolf, A. B. Bhatia, P. C. Clemmow, D. Gabor, A. R. Stokes, A. M. Taylor, P. A. Wayman, W. L. Wilcock, *Principles of Optics*, Cambridge University Press, 1999.
[38] D. Momeni Pakdehi, P. Schädlich, T. T. N. Nguyen, A. A. Zakharov, S. Wundrack, E. Najafidehaghani, F. Speck, K. Pierz, T. Seyller, C. Tegenkamp, H. W. Schumacher, *Adv Funct Mater* 2020, *30*.

## Supporting Information

### Table of contents:

### S1. Near-field imaging of fabricated 2D-Ag/EG plasmonic nano-disks

The studied 2D-Ag/EG plasmonic nanostructures on the SiC substrate are fabricated by the following process. The EG is firstly grown on SiC substrate by silicon sublimation on the Si face of 6H-SiC (0001). Then, the intercalation process happens through thermal evaporation: vaporized Ag atoms diffuse in between the EG layer and SiC substrate and form a crystalline 2D-Ag layer. Afterwards, a mask with designed plasmonic shape will be formed by e-beam lithography and the areas outside the masked region will be removed by the chemical etching process. Finally, we will obtain 2D-Ag/EG plasmonic nanostructures (photonic dots) on the SiC substrate.

The EG sample, formed upon epitaxial growth process, may include a partial monolayer EG on top of the buffer graphene layer. The 2D-Ag layer will form between the buffer graphene layer and SiC substrate during the intercalation process. Therefore, the Ag layer is encapsulated by the monolayer graphene and partial two-layer graphene in different areas. The substrate outside the plasmonic nanostructures is expected to be bare SiC.

The area we studied consists of a series of disk-shape 2D-Ag/EG nanostructures, where we performed near-field imaging. Representative maps at the excitation wavelength 968 $cm^{-1}$ are shown: sSNOM/optical maps, Figure S1c-j, were taken simultaneously with the mechanical/AFM channels, Figure S1a-b. From AFM topography and (mechanical) phase image, we can distinguish the terrace step edges of SiC substrate as well as the shape of the nano-disks. The size of the nano-disks is below 1 μm, increasing from left to right.

The “snake-like” morphologies within the nano-disks have 0.7 nm height variation that are consistent with the material composition of incomplete two-layer EG. In the course of metal interaction, monolayer or bilayer of quasi-free-standing graphene are formed from a buffer or a buffer plus a monolayer EG. In case the initial EG contains an incomplete monolayer, the resulting graphene composition after Ag intercalation will be less than 2 and greater than 1 layer, thus forming irregular partial bilayer structure.

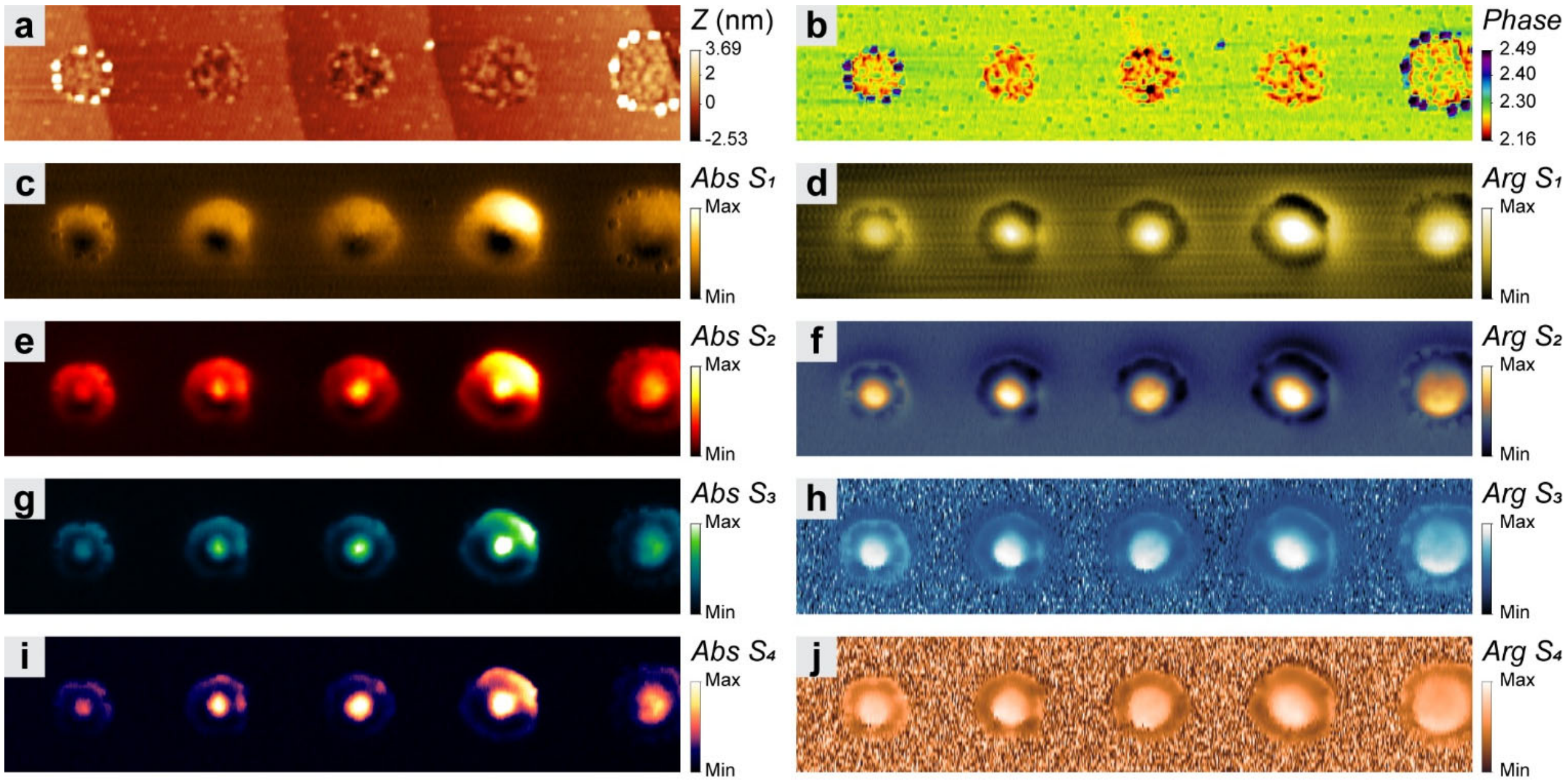

**Figure S1.** Multichannel sSNOM maps of 2D-Ag/EG plasmonic nano-disks. Scale bar 500 nm.

Figure S1c-j shows the simultaneously obtained multichannel sSNOM optical maps, i.e., different harmonic demodulated (1st to 4th) sSNOM signals: the optical amplitude (Abs($S_1$) to Abs($S_4$)), and the optical phase (Arg($S_1$) to Arg($S_1$)) of a series of nano-disks. Upon visual inspection, we notice that the region inside the nano-disk shows a higher optical amplitude signal compared to the outside region (substrate). Also, comparing the phase and amplitude maps, we trace a wide-belt region around the circumference of the nano-disk: it has lower signal, though much higher than of the substrate in all optical amplitude maps, and shows a clear contrast compared to both the region inside the nano-disk and the substrate in all optical phase maps. Need to mention that the optical response we observed in these fabricated nano-disks does not follow nor has a substantial dependence on the surface morphology of the sample, i.e., the near-field signal is not substantially affected by the existence of impurities/residues, induced during the fabrication/etching process. Obviously, there is no clear trace of wave patterns formed within the region of nano-disk nor in the outside region of the nano-disk, which is easily explained by having about four order of magnitude difference in scale between the disk size and laser wavelength.

Hyperspectral sSNOM mapping was performed with excitation wavelength 963-1040 $cm^{-1}$ of the region of interest, i.e., the third disk counted from left in Figure S1a. The representative sSNOM Abs ($S_3$) and Arg ($S_3$) maps are shown in Table S1 and SI video file MP4 S1.

**Table S1.** Hyperspectral sSNOM maps of the region of interest.

| | | | | | |
|---|---|---|---|---|---|
| 963 $cm^{-1}$ | 964 $cm^{-1}$ | 965 $cm^{-1}$ | 966 $cm^{-1}$ | 967 $cm^{-1}$ | 968 $cm^{-1}$ |
| 969 $cm^{-1}$ | 970 $cm^{-1}$ | 971 $cm^{-1}$ | 972 $cm^{-1}$ | 973 $cm^{-1}$ | 974 $cm^{-1}$ |
| 975 $cm^{-1}$ | 976 $cm^{-1}$ | 977 $cm^{-1}$ | 978 $cm^{-1}$ | 979 $cm^{-1}$ | 980 $cm^{-1}$ |
| 981 $cm^{-1}$ | 982 $cm^{-1}$ | 983 $cm^{-1}$ | 984 $cm^{-1}$ | 985 $cm^{-1}$ | 986 $cm^{-1}$ |
| 987 $cm^{-1}$ | 988 $cm^{-1}$ | 989 $cm^{-1}$ | 990 $cm^{-1}$ | 991 $cm^{-1}$ | 992 $cm^{-1}$ |
| 993 $cm^{-1}$ | 994 $cm^{-1}$ | 995 $cm^{-1}$ | 996 $cm^{-1}$ | 997 $cm^{-1}$ | 998 $cm^{-1}$ |
| 999 $cm^{-1}$ | 1000 $cm^{-1}$ | 1001 $cm^{-1}$ | 1002 $cm^{-1}$ | 1003 $cm^{-1}$ | 1004 $cm^{-1}$ |
| 1005 $cm^{-1}$ | 1006 $cm^{-1}$ | 1007 $cm^{-1}$ | 1008 $cm^{-1}$ | 1009 $cm^{-1}$ | 1010 $cm^{-1}$ |
| 1011 $cm^{-1}$ | 1012 $cm^{-1}$ | 1013 $cm^{-1}$ | 1014 $cm^{-1}$ | 1015 $cm^{-1}$ | 1016 $cm^{-1}$ |
| 1017 $cm^{-1}$ | 1018 $cm^{-1}$ | 1019 $cm^{-1}$ | 1020 $cm^{-1}$ | 1021 $cm^{-1}$ | 1022 $cm^{-1}$ |
| 1023 $cm^{-1}$ | 1024 $cm^{-1}$ | 1025 $cm^{-1}$ | 1026 $cm^{-1}$ | 1027 $cm^{-1}$ | 1028 $cm^{-1}$ |
| 1029 $cm^{-1}$ | 1030 $cm^{-1}$ | 1031 $cm^{-1}$ | 1032 $cm^{-1}$ | 1033 $cm^{-1}$ | 1034 $cm^{-1}$ |
| 1035 $cm^{-1}$ | 1036 $cm^{-1}$ | 1037 $cm^{-1}$ | 1038 $cm^{-1}$ | 1039 $cm^{-1}$ | 1040 $cm^{-1}$ |

**S2. Correlated arc shape in Argand space: indication and verification of surface polariton wave**

The near-field optical signals we obtained from sSNOM measurements are complex valued: $S_3 = Ae^{i\varphi}$, where the optical amplitude is $A = Abs(S_3)$ and the phase $\varphi = Arg(S_3)$. One can represent the sSNOM complex valued response by the real ($Re\, S_3 = Acos\varphi$) and imaginary part ($Im\, S_3 = Asin\varphi$).

Figure S2a is the cross-correlation plot between the imaginary and real parts, where the gray scatter dots are the $S_3$ data, taken at excitation wavelength 1030 $cm^{-1}$, from the whole region of interest. This panel reproduces the Figure 2e in the main text. Figure S2b-c are obtained by the Cartesian-to-polar coordinate transformation of the third demodulated harmonic optical amplitude, Abs($S_3$), and phase, Abs($S_3$), maps of the region of interest from raw sSNOM data. The blue scatter dots in panel (a) represent the data from a single slice (with 4 pixels in $\theta$ direction) of the polar maps, highlighted by green rectangles in Figure S2b, c. Two adjoint arcs can be clearly traced (blue scatter dots). The shape of an arc in the complex plane corresponds to an eikonal wave with the constant amplitude and variable phase of a phasor of the sSNOM signal, indicating the existence of a propagating surface polariton (SP) wave.

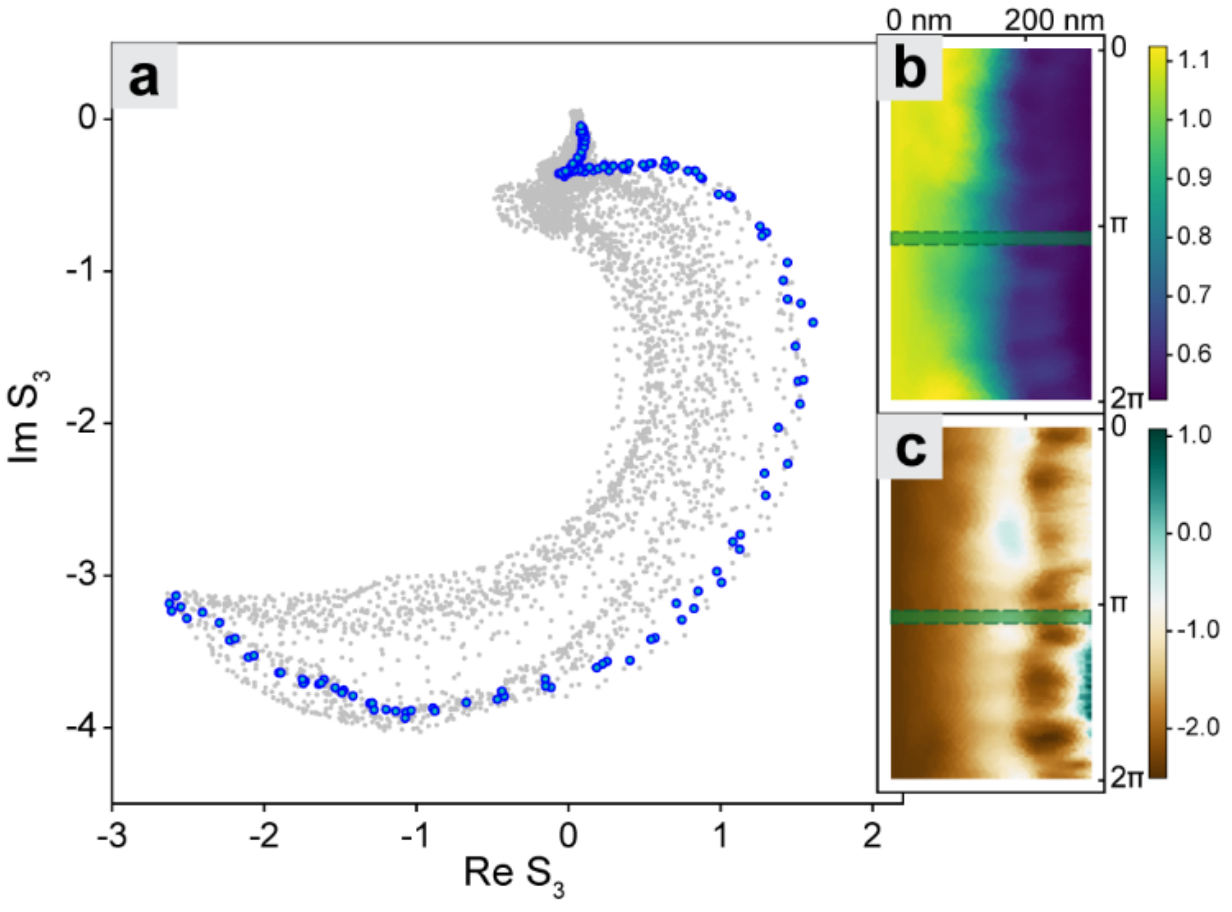

**Figure S2.** Correlated arc shape in Argand space.

To further verify the existence of radially propagating surface polariton wave, we focus on a single arc in Argand space and trace the rate of phase variation along the arc with respect to spatial displacement of their corresponding locations in the real space map. To determine its propagation direction, in Figure S3a we choose in Argand space a sequence of segments (marked by the rainbow colors) with the fixed amplitude (same as the radial coordinate for the spokes from the center of the arc). The Figure S3b shows that the trajectory, from red to magenta, moves from the disk center to its edge in the real space. For an outgoing radial wave we expect the counterclockwise evolution of the phasor in Argand space (the clockwise evolution would correspond to inward radial wave). Indeed, taking a sequence of trajectory points/small regions in the real space displaced in radial direction, as shown in Figure S3d, their counterparts in Argand space follow the counterclockwise arc, plotted in Figure S3c, which supports our assumption.

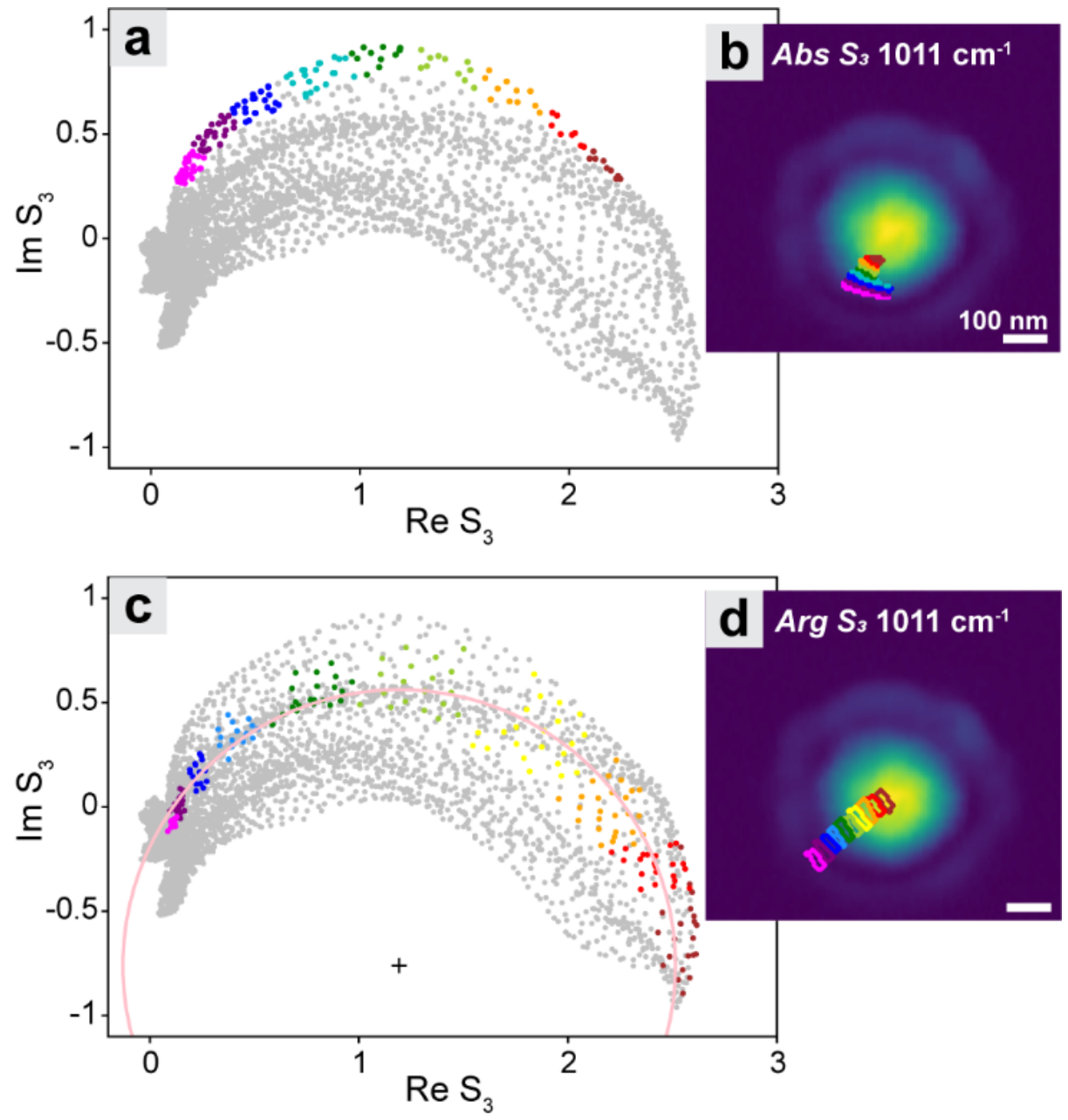

**Figure S3.** Correlating trace of phase in Argand space vs. polariton wave in real space. (a) Selected clusters in Argand space with variation in phase but nearly equal amplitude. (b) The corresponding data points in real space form a radial wedge of the outgoing radial wave. (c) Data points in Argand space from the sectors selected in (d). All the selected data points are fitted to a circle (pink curve) with the center marked as a black cross. (d) Areas selected in the real space along a single radial direction.

**S3. Referencing point: background screening by different material compositions**

The sSNOM signal we obtained consists of (1) a propagating wave component, stemming from the near-field response of propagating surface polaritons, possibly confined in the photonic dot, and (2) the background signal or non-propagating component, of which the majority is due to the dielectric screening by the spatially uniform sample material. Therefore, to quantitatively measure the eikonal wave properties, such as the wave amplitude and wavevector of the surface polaritons, proper referencing needs to be performed first.

In our new method, the referencing is done by fitting the eikonal arc segment with a circle in the Argand space, where the coordinates of the center of the circle correspond to the complex valued background screening by the material, while the radius of the circle is the true magnitude of the eikonal wave and the phase velocity (the rate of change of the true phase) reveals the propagation constant of the wave.

Besides the larger arc (represented by red scatter dots), we notice a "tail swirl" (represented by green scatter dots), a small segment in the Argand plane which clearly falls out of the single arc sequence in Figure S4a. Clearly, this segment possesses a different referencing point compared to the red arc, as well as a different magnitude (radius of the circle) of the wave. We fitted this segment of the arc (green scatter dots) in Figure S4a. The smaller radius of arc corresponds to substantially smaller magnitude of the radial wave in this spatial location. Specifically, the green dots data come from the belt region near the circumference of the nano-disk. We assume it is composed of oxidized 2D-Ag. We note that the eikonal waves confined within the belt and the nano-disk propagate in the same (outward radial) direction (which can be seen from the counterclockwise evolution of the phasor). The smaller magnitude indicates the weak "leak" of the eikonal wave from bare 2D-Ag center to oxide belt, which results in substantial drop of the wave magnitude (and therefore, the EM field energy).

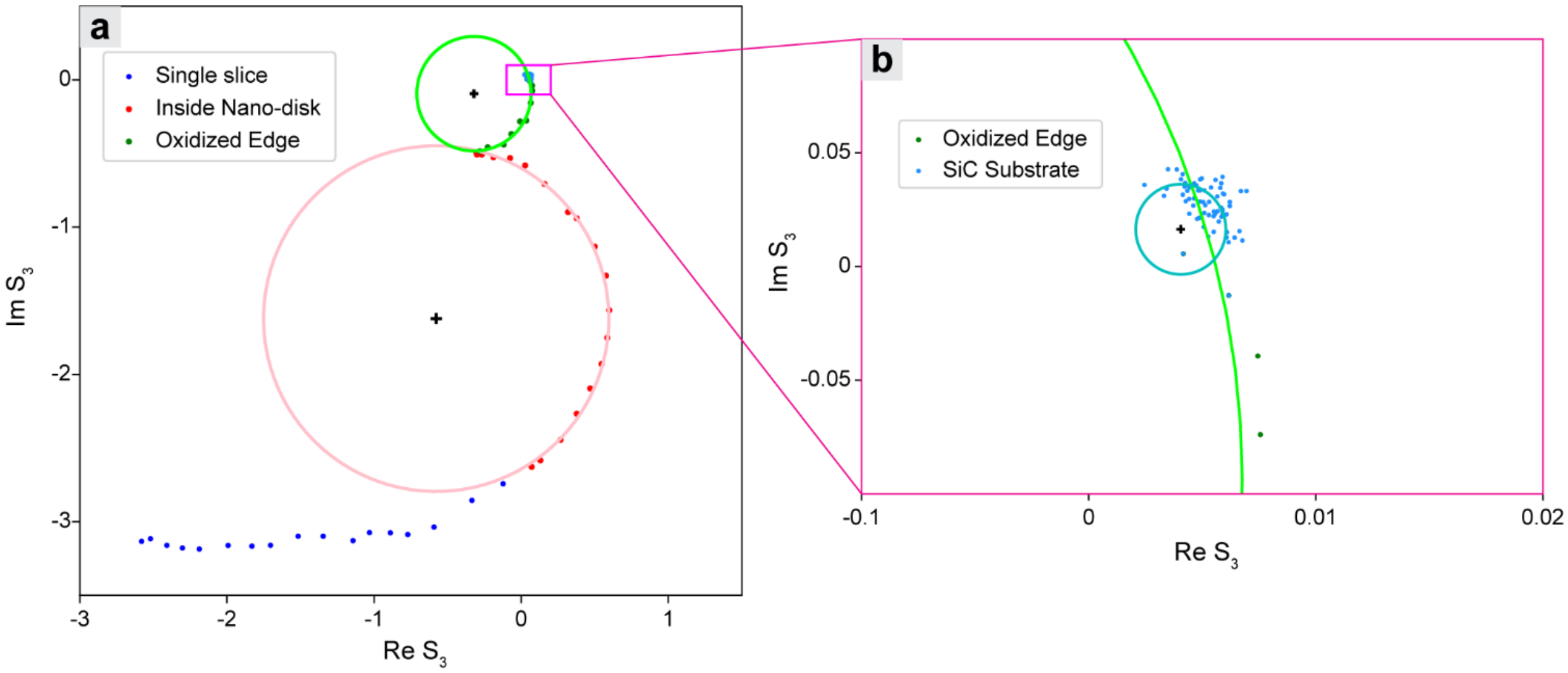

**Figure S4.** Fitting eikonal wave arcs for 3 different regions. (a) Main part of the Argand plot shows the arc due to the bare Ag material in the central region of the photonics quantum dot (pink) and the oxide belt (green) – both encapsulated by the EG layer; (b) a magnified view of the last segment of the phasor curve shows the arc for the bare SiC substrate (cyan) (part of the oxide belt data is also shown).

Besides these two arcs, corresponding to bare Ag and oxide belt, we notice another short arc segment at the end of "tail swirl", which can be seen better in the zoomed-in Argand plot (Figure S4b). We can attribute these data points to the region outside of nano-disk, that is to the eikonal wave on the SiC substrate. The radius of bare SiC arc (cyan) in Figure S4b is far smaller than the radius of light green circle, thus, confirming much smaller true magnitude of the wave outside of the quantum dot and proving our statement on the complete confinement od SP within the nano-disks.

We emphasize that these different arc segments also correspond to different reference signals, which means that the different optical background needs to be subtracted to obtain the true phase of eikonal wave. Since the referencing point (background) is due to the dielectric screening of the tip dipole (in the tip-sample image dipole model), it indicates the different material composition in these regions.

The coordinates of these referencing points (the backgrounds) vary with respect to the excitation frequency. This is due to the material optical response depends on the excitation (Figure 4a). Both the region inside the nano-disk (SiC/bare 2D-Ag/EG) and in the belt (SiC/oxidized 2D-Ag/EG) show dispersion, typical for a response near the frequency pole. By comparing the pole position to the SiC response function we deduce that the non-monotonic behavior is due to the optical phonon of the latter.

### S4. Dispersion relation of polaritons

In order to determine dispersion of the eikonal wave, the real-space image for each excitation wavelength has been transformed from cartesian into polar coordinates. Then, the angular slices with fixed width (data for a narrow region of angles near the $\theta$ value) are fitted to the arc segments: the center (reference point) and the radius (true wave magnitude) in Argand space are defined (Figure 3). The evolution of the true phase is determined vs. the radial coordinate, $r$, which is limited to the geometrical region of the nano-disk, approximately $r$<260 nm. The overall arc length in the Argand space, $\phi_{tot} = \phi(r_{max}) - \phi(r_o)$, for each slice is computed and related to the corresponding trajectory length in the real space, $L = r_{max} - r_o$, which allows to define the mean eikonal wavelength of the surface polariton within this arc segment:

$$\Lambda = \frac{2\pi L}{\phi_{tot}}$$

The SP eikonal wavelength is defined for each excitation frequency to produce the dispersion relation.

Consequently, we can plot excitation frequency vs. average polariton propagation constant, obtained as: $k_{tot} = \frac{\phi_{tot}}{L}$. Figure 4d shows the dispersion relation obtained within the nano-disk (i.e., using the red scatter dots from the major arc segment in Argand space) and averaged over the angular coordinates. The red markers represent the mean values of wavevectors for each excitation wavelength. We clearly see two branches of surface phonon polariton originated from SiC, denoting the Reststrahlen band of SiC optical phonon. This dispersion relation obtained from the overall arc length $\phi_{tot}$ is an integral property of the disk with the size $L$. Notably, the arc length is not just a linear function of $L$ (see Supplementary Section S5).

**S5. Characteristics of eikonal waves**

Since we have already established radial propagation direction of the polariton wave, we can proceed with selecting the appropriate segments in the real-space map, starting from the center towards the edge of the nano-disk, and find the corresponding clusters in Argand space.

Notice that in Figure S3c, the clusters in the Argand space are spread less tight for the location near the edge of the nano-disk, while they distribute more widely for the clusters near the center, e.g., data points from the magenta cluster show less fluctuation of the eikonal wave magnitude, $M$, than those from the dark red cluster. Thus, eikonal wave is better defined for the magenta segment, compared to the red segment. Partially, this may be due to the transformation to radial coordinates must reduce the angle/radial pixel accuracy near the coordinate origin. Additionally, the arc length of the phase varies from segment to segment in the Argand space, despite the same size of each segment in the real space, indicating the nonlinear change of the true phase with the propagation distance, i.e., the locally varied propagation constant of the eikonal wave.

To further examine the behavior of the surface polariton waves, we can transform the Argand plot (in polar coordinates) into the true magnitude and true phase coordinates after fitting the scatter dots to a circle and subtracting the reference center, for both red and green arcs. Figure S5b shows the true magnitude and angle from the data points along a specific radial direction in real space, where we can clearly observe the damping of the eikonal wave amplitude towards the edge. However, since the damping parameter is small we will neglect this for the short distance from the disk center to the edge when fitting the radial waves for the rest of the work.

Figure S5a shows the representative plot in Argand space for a specific angular slice in real space at excitation wavelength 994 $cm^{-1}$. The data points (marked with alternating color) are taken from the small radial boxes of the same size, i.e., for the constant increment of radial displacement in the real space. Using this construction, one can compare corresponding arc length in the Argand space as a function of radial distance from the disk center. The length of the arc segments near the middle of the range (highlighted by pink rectangle in Figure S5b) is larger compared to the ends of the arc, which can be better observed in the plot of the true magnitude and angle (Figure S5b). This hints that the propagation constant of the surface phonon polariton may be a function of coordinates. The true phase gradient (derivative along the correlated arc in the Argand space with respect the trajectory displacement in the real space) is computed for each point/small domain on the sSNOM map, as shown in next section.

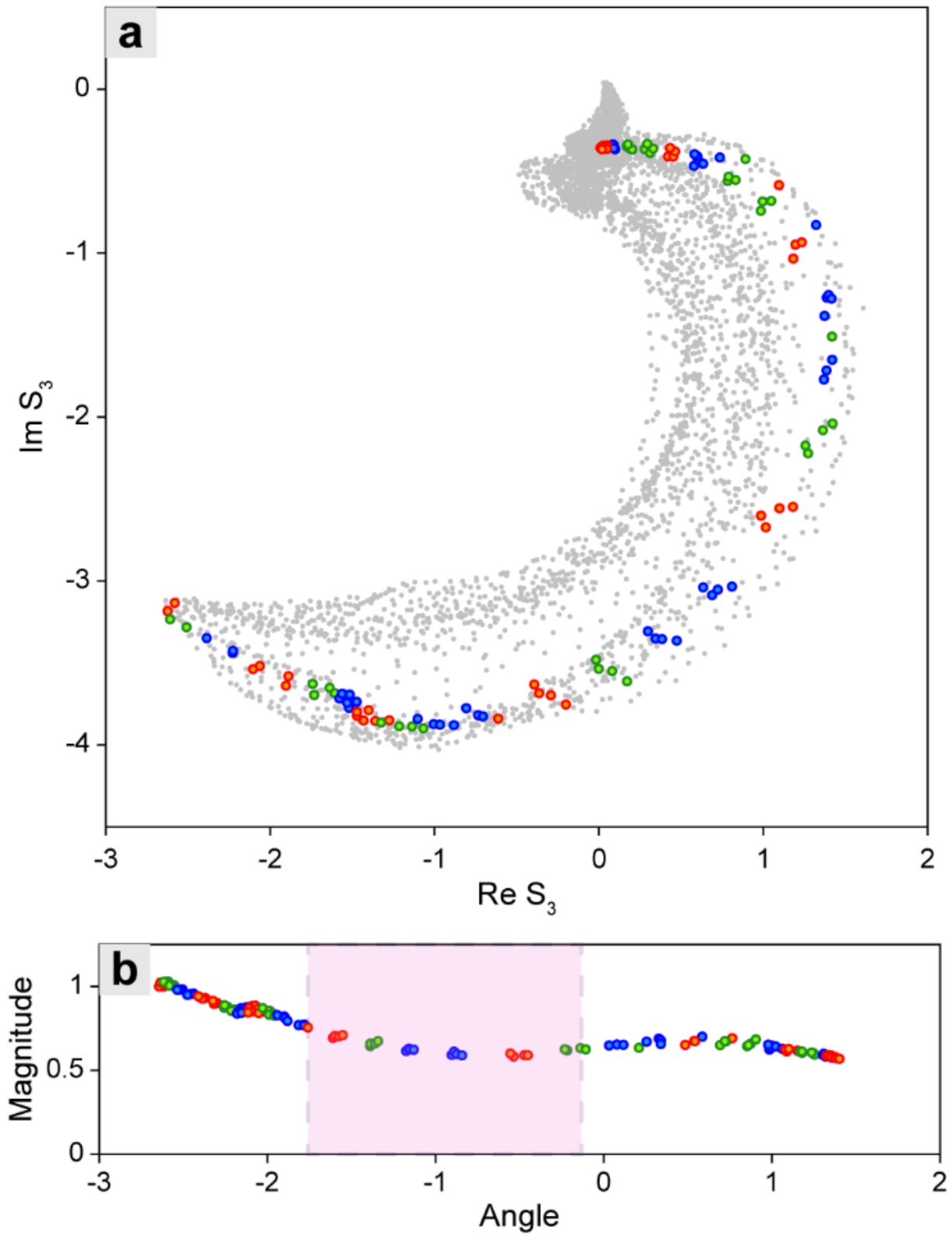

**Figure S5.** Arc fitting: the phase velocity as a function of the phase, proportional to the propagation length, $L$.

## S6. Evaluation of eikonal model

Figure S6 shows the representative gradient maps of true magnitude and true phase with respect to the $\theta$ and $r$ at the excitation wavelength 994 $cm^{-1}$. As we can see, the values of the angular gradients: $\frac{\partial M}{\partial \theta}$ (Figure S6a) and $\frac{\partial \phi}{\partial \theta}$ (Figure S6c) are close to zero, indicating a little evolution in angular direction, which is consistent with the wavefront of the eikonal wave in the real space being nearly circular. The map of $\frac{\partial M}{\partial r}$ (Figure S6b) evaluates the deviation of the sSNOM data from our assumption of an ideal eikonal wave, i.e., it shows how close is the arc shape to the circle. Including distance dependence of the wave magnitude, the phasor can be represented as following:

$$A_o e^{iks+log\frac{M(s)}{M_o}}$$

where $M_o$ is the extracted magnitude in Argand space of a data point taken as the wave origin (the center of nano-disk in real space in this case). Naturally, the second term in the exponent gives a logarithmic derivative component to the calculated propagation constant. Figure S5b shows that the true magnitude may deflect from a constant value for some slices (this effect is beyond the oxide belt or SiC region, discussed earlier). However, the variation of the gradient $\frac{\partial A}{\partial r}$ (observed in the map in Figure S6b) is small compared to true phase term.

Within these approximations, the map of $\frac{\partial \phi}{\partial r}$ (Figure S6d) shows the local values of the propagation constant of the eikonal wave.

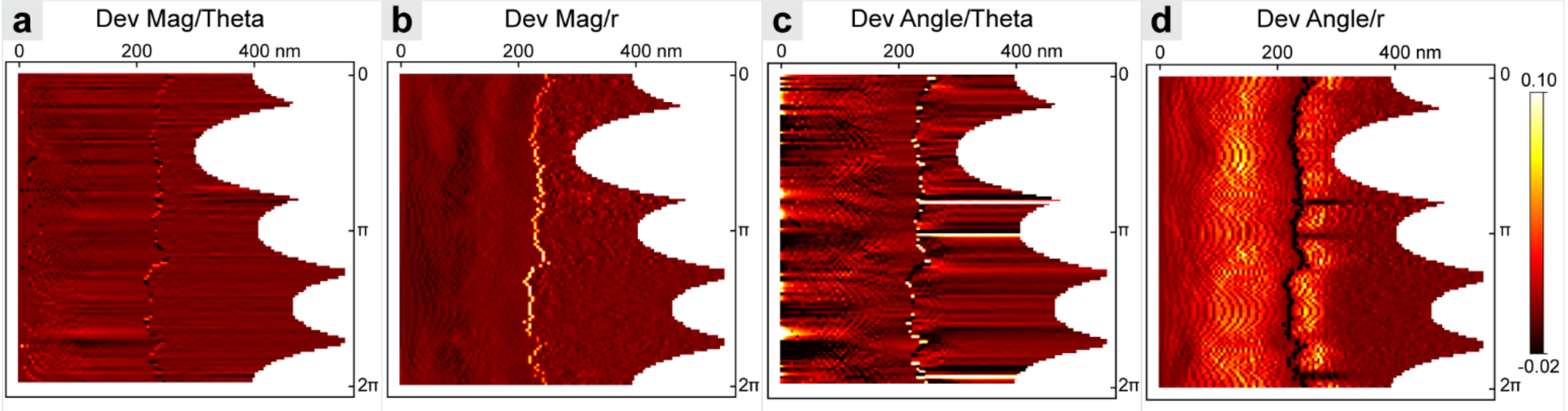

**Figure S6.** Evaluation of an accuracy of the radial eikonal wave model: the maps of all components of the phase velocity gradient. The map of derivative of magnitudes with respect to (a) angular and (b) radial direction. The map of derivative of angular coordinates with respect to (c) angular and (d) radial direction.

**S7. Validation of eikonal model results**

Figure S7 shows the AFM and sSNOM maps of 5 more photonic Quantum Dots (pQDs), along with the one used for main text analysis (disk 13 in yellow rectange). All disks have different diameters, though made of the same material. Obviously, all pQDs demonstrate a similar ultr-strong polariton confinement, which is further supported by taking the data for the eikonal analysis.

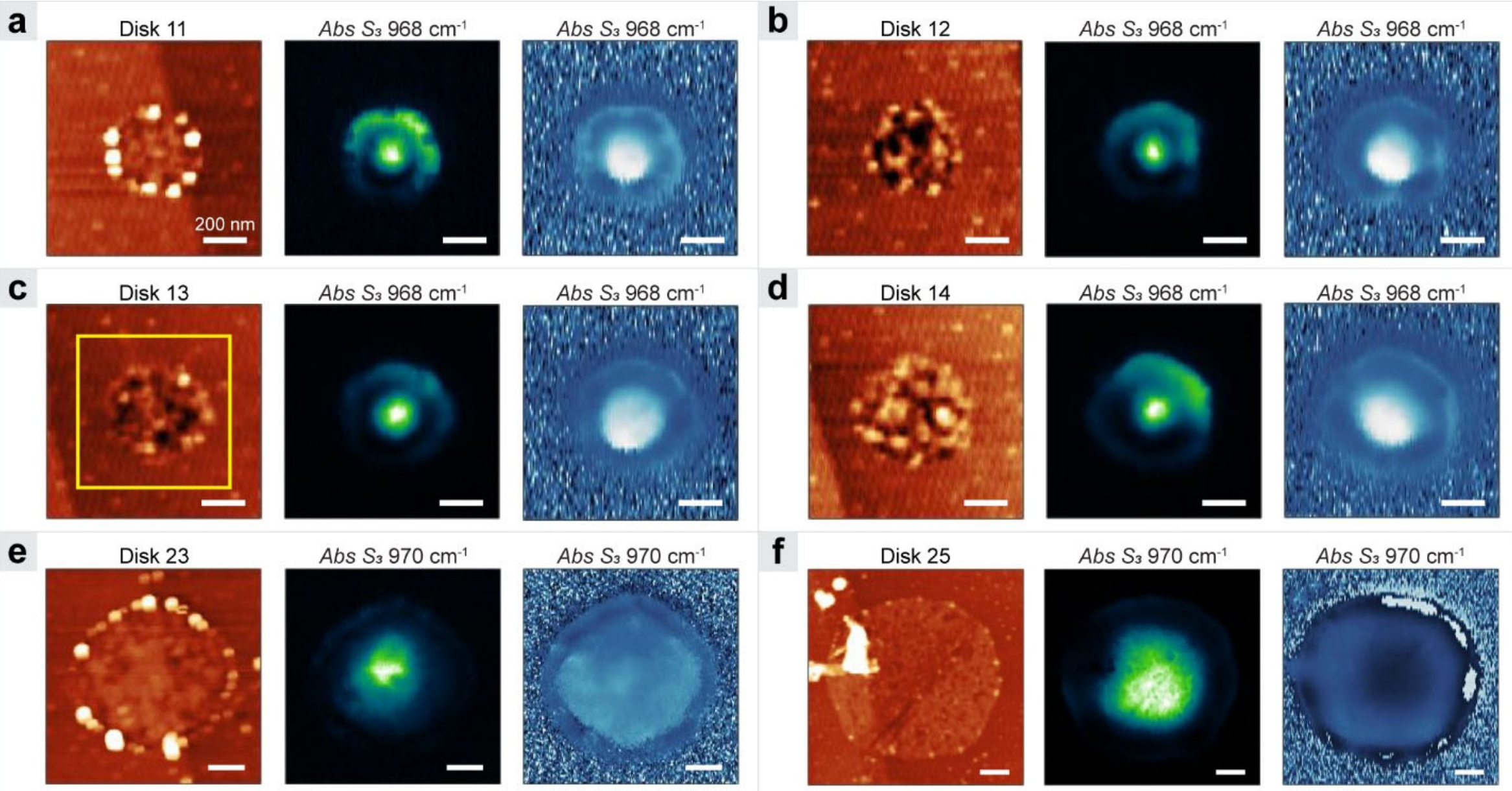

**Figure S7.** sSNOM maps of 2D-Ag/EG plasmonic nano-disks with different sizes. AFM topography image, sSNOM Abs($S_3$), and Arg($S_3$) maps at excitation wavelength 968 cm$^{-1}$ of nano-disk 11-14 (a-d); AFM topography image, sSNOM Abs($S_3$), and Arg($S_3$) maps at excitation wavelength 970 cm$^{-1}$ of nano-disk numbered 23, 25 (e, f). Scale bars are 200 nm. The nano-disk studied in the main text is marked with yellow square in panel (c).

The sSNOM maps from Figure S7 were processed by the same eikonal wave analysis as in main text and resulting propagation constants are plotted vs. the measured pQD diameter (inverse of) in Figure S8. Overall, a clear correlation of the propagation constants for pQDs of a similar diameter is seen, with very little variability, along with some monotonic dependence on the pQD size. Notably, the periodicity (wavelength) is smaller than the disk diameter for large enough disks.

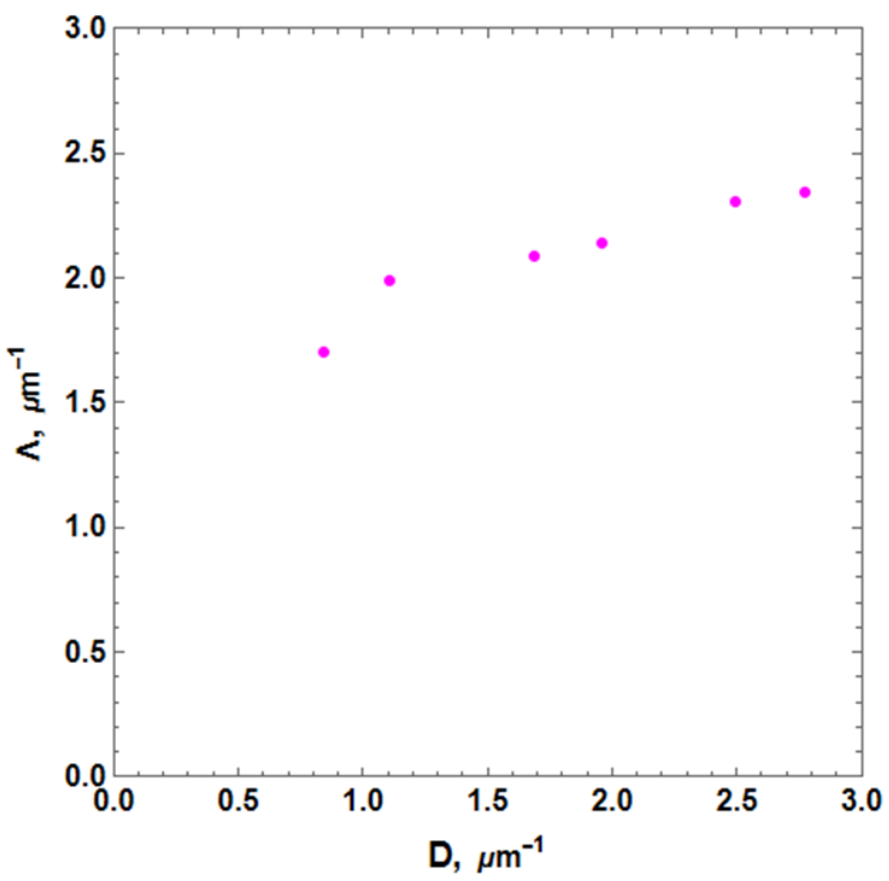

**Figure S8.** Propagation constants for series of pQDs of different diameters. Measured inverse wavelength (propagation constant) for pQDs #11-14 and #23, #25 from Figure S7 obtained from the raw sSNOM Abs($S_3$) and Arg($S_3$) maps by the eikonal wave analysis.

The sSNOM mapping data might potentially contain measurements artifacts, mostly due to a mechanical AFM tip motion instability when moving over a large surface impurity. Since the surface morphology of the pQDs is not perfect (due to the fabrication process), we took additional care to exclude cases that might produce ambiguous eikonal wave analysis. Figure S9 provides examples of "negligible" and "non-negligible" effect of surface impurities on the optical sSNOM signal. Panel (a) shows AFM maps of 5 pQDs, including the one which is used in main text analysis. Panels (b-e) show 4 representative profiles taken along the arrows labeled with the corresponding numbers in the panel (a). In all 4 examples, neither Abs($S_3$) nor Arg($S_3$) signals (green and blue lines) show substantial correlation with the AFM profile (brown line). On contrary, last example in panel (f) shows significant correlation between AFM and Arg($S_3$) channels and should be excluded from the analysis. Note that the height of the impurity in this case exceeds 6 nm, in contrast to 1-2 nm in other cases.

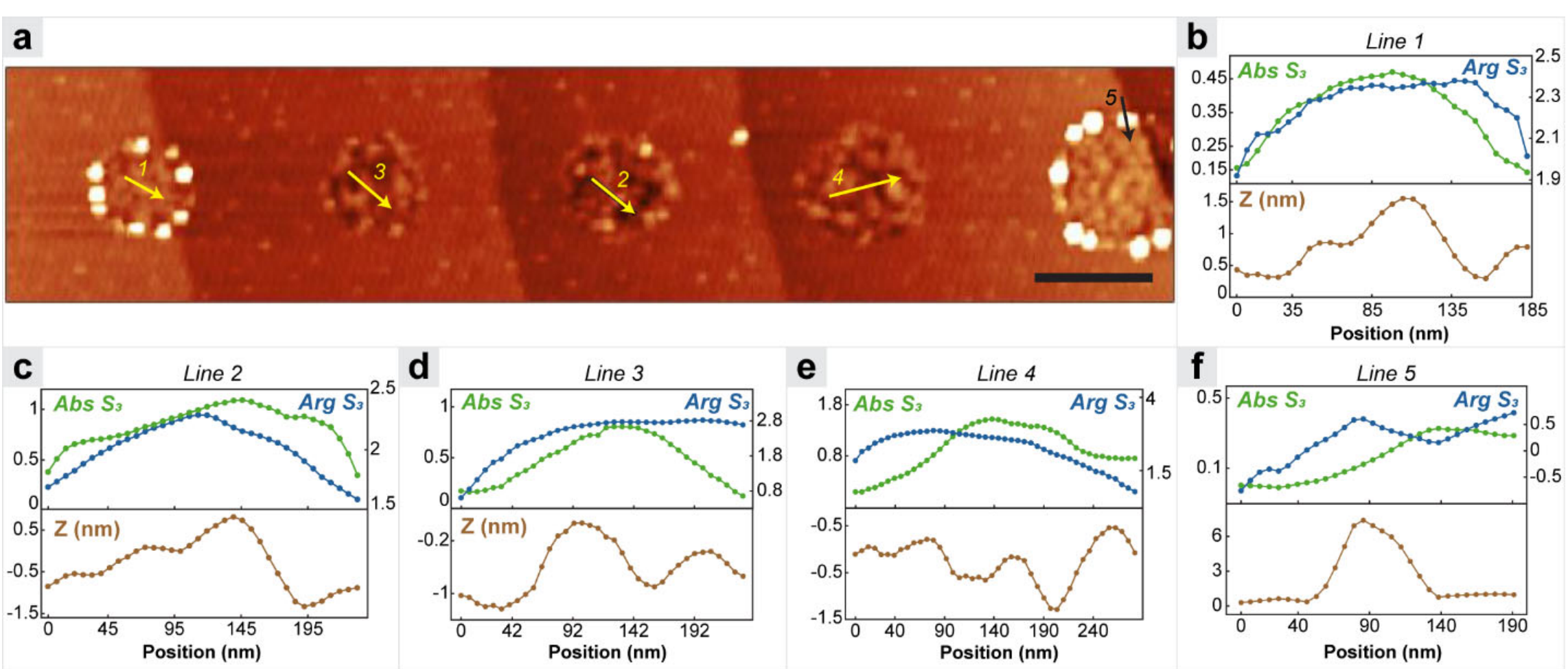

**Figure S9.** Line profiles of sSNOM signals compared to the AFM topography. (a) AFM topography image of nano-disk 11-15. The locations of the corresponding line profiles shown in panel (a-f) are indicated by arrows. The scale bar is 500 nm. (b-e) Line profiles of the sSNOM Abs($S_3$) and Arg($S_3$) (top graph), compared to the AFM Z height (bottom graph) of lines 1-4, where the sSNOM signals show no correlation with the AFM Z height. (f) Line profiles of the sSNOM Abs($S_3$) and Arg($S_3$) (top graph), compared to the AFM Z height (bottom graph) of line 5, where the sSNOM Arg($S_3$) shows some correlation with the AFM morphology due to an impurity with a large height profile.

**S8. Validation of eikonal model analysis with hBN/Gr/hBN polaritons**

To provide additional evidence for legitimacy of novel eikonal analysis we present here a hyperspectral mapping series on another polaritonic material (unrelated to main study on SiC polariton confinement). In this SI section we provide analysis of polariton modes' dispersion in hBN/Graphene/hBN sandwich, transferred onto Si/$SiO_2$ substrate. The sample was prepared with standard transfer methods: the bottom inset in Figure S10a shows schematically the Si-substrate with 285 nm oxide layer, (thick) bottom layer of hBN, followed by monolayer graphene (MLG) and another (thin) top layer of hBN, corresponding to the cross-section along dashed line in (a). The structure was thouroughly characterized by Raman microscopy – Figure S10e shows Raman signature bands for each layer of the sandwich material. We focus next on the edge area, where the (conformal) film of top-hBN/MLG falls of the cliff of (thick) bottom layer of hBN to the bare $SiO_2$/Si substrate (as moving from left to right along the dashed red line in Figure S10a). In this area we expect strong reflection of polaritonic waves from the edge.

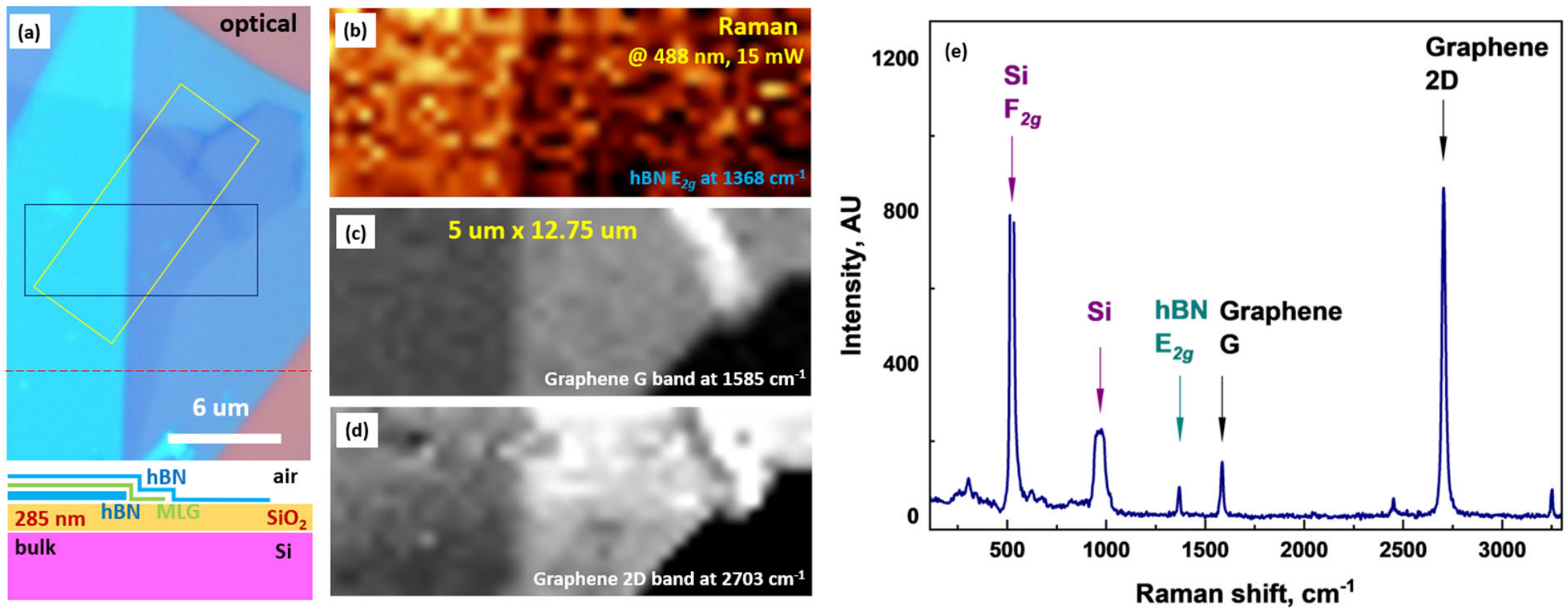

**Figure S10.** Raman characterization of the sample SH4. (a) Optical image of the sample; (bottom inset) the cartoon schematics of the transferred layers as appear along the chosen cross-section (red dashed line) in the micrograph. (b-d) Representative Raman maps to identify graphene (Gr) and hBN layers. The maps are taken over the same area 12.75um x 5um, shown as a black rectangle in (a). The maps are shown for position of characteristic peaks of (b) $E_{2g}$ band of hBN, (c) G and (d) 2D bands of GR, as selected from (e) the Raman spectrum. These peaks were clearly identified in the range 250-3600 $cm^{-1}$. Excitation wavelength was 488 nm.

Figure S11 shows the sSNOM maps of top-hBN/MLG/b-hBN/$SiO_2$/Si sandwich material, where the hyperspectral series SH4 (11 single laser frequency maps) were taken. Abs and Arg of $S_2$ sSNOM signal at 1410 $cm^{-1}$ are presented in Figure S11b,c. Physical structure is well resolved, including the edge between the left region of full sandwich structure (t-hBN/MLG/b-hBN/$SiO_2$/Si) and the right region with missing bottom hBN layer (t-hBN/MLG/$SiO_2$/Si). Graphene folds are also clearly resolved in the dark area in Figure S11b.

More detailed mapping was done over the area next to this edge, as shown in Figure S11d. Here Abs $S_3$ sSNOM signal is used, the same which is analyzed in the eikonal wave model next. An "ordinary" method of obtaining the polariton dispersion is to count the distance between the maxima of the wave pattern – those are indicated by arrows in panel (d). Notably, the distance between the peaks is not necessarily uniform which makes a large uncertainty for polariton dispersion (to be discussed below). For the sake of convenience of further data analysis, we performed a numerical transformation for hyperspectral map series to make the edge line approximately vertical (since we use raw data, without any interpolation, an exact alignment is limited by the width of the pixel in horizontal direction, which was chosen to be 8 nm in our case). The pre-aligned data is shown in panel (e); the map corresponds to the same height area as in panel (d) with the rugged edge (not shown), as needed for the alignment.

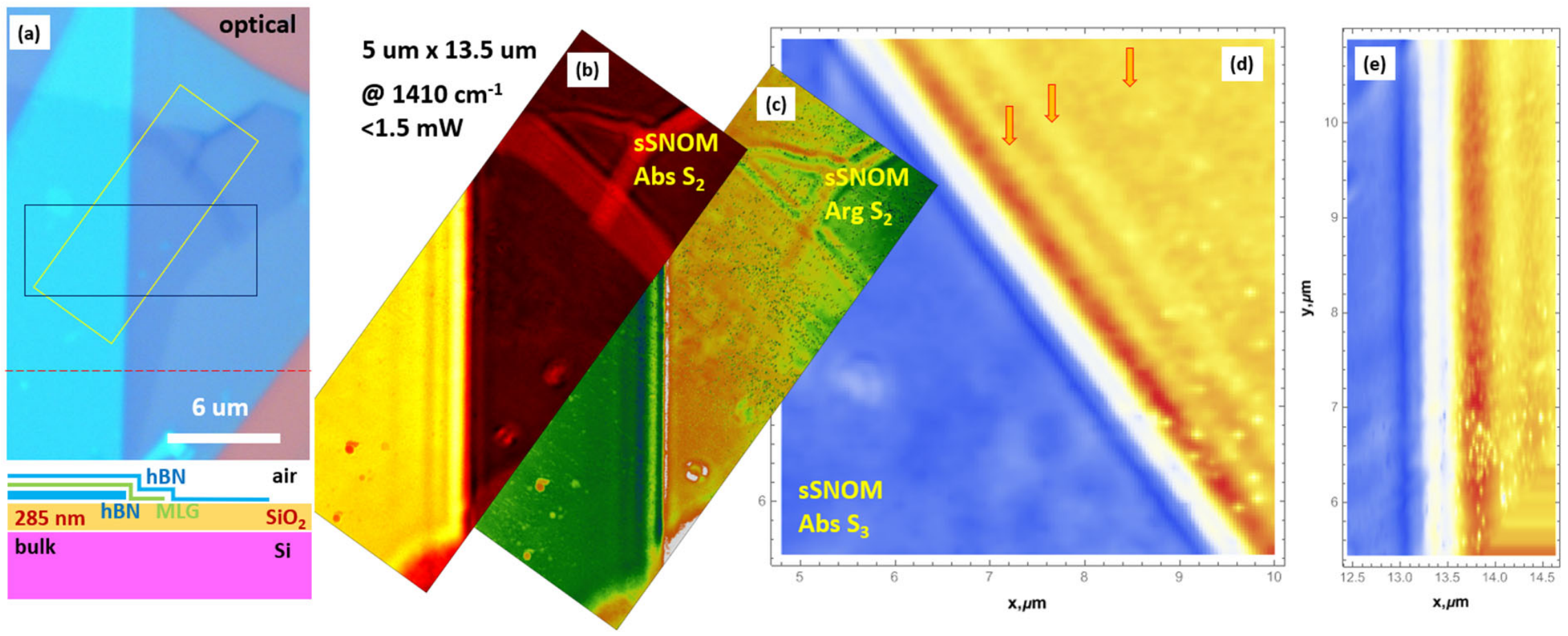

**Figure S11.** sSNOM mapping of the hBN polaritons. (a) same optical micrograph as in Figure S10. (b-c) Maps of the area shown by yellow rectangle in (a): O2A and O2P channels are presented. sSNOM contrast clearly distinguishes the edge of thick bottom hBN layer in a good correlation with the Raman maps (cf. Fig. S10b); in (b) sSNOM also reflects graphene folds in the top of the map (cf. Fig. S10c). (d-e) Maps of O3A at 1410 $cm^{-1}$. Arrows in (d) indicate maxima of polaritonic wave signal to be used for manual detection of polariton wavelength (note a non-uniform separation between peaks). (e) sSNOM aligned map as obtained by numerical transformation of (d) to keep edge along vertical direction.

In contrast to “ordinary” wave-peak counting approach, which is ambigous, we apply fully numerical analysis via the eikonal wave. We derive the wavevector of polariton by detecting the polaritonic eikonal phase derivative, $-i\,\partial\phi/\,\partial x$, which we compute moving across the edge boundary. The result is shown in Figure S12a. Importantly, the wavevector in the eikonal model is not global but rather a local quantity (depends on the wave location). While this gives an advantage to materials characterization, it also requires a careful analysis. Specifically, in case of polariton waves reflected from the edge, the propagation constant has an opposite sign to the left and right side of the interface. As a result, the phase of the wave incident from the left should be decreasing (with negative derivative), and then, after edge point, increasing (with positive derivative). It is well known that the sSNOM signal is coming already pre-processed from the Neaspec software. As one can see in Figure S12e, the sign of the phase is exactly opposite, which is a typical artifact of the harmonics’ demodulation. Nevertheless, by knowing the physics of the wave scattering process, one can reproduce the right direction of wave propagation. In our case, the trajectory in the Argand space starts from top (near smaller black fitting circle) and continues downwards (to blue circle). The points in this Argand plot are from the whole map shown in Figure S12d and clearly form two “kissing” arcs for waves reflected off the edge. The central positions of the fitting circles are not the same, which allows one to determine the shift of the sSNOM background signal when moving off the bottom-hBN layer.

Since not only the phase derivative changes the sign but also the background signal (subtracted center position of the fitting circle) is different on two sides of the edge, the curves like in the Figure S12a are different for fitting on either side of the interface (here we only get the values on the right). However, one can see a good convergence of the wavevector to the value derived from the manual wavelength count (depicted by the blue horizontal gridline).

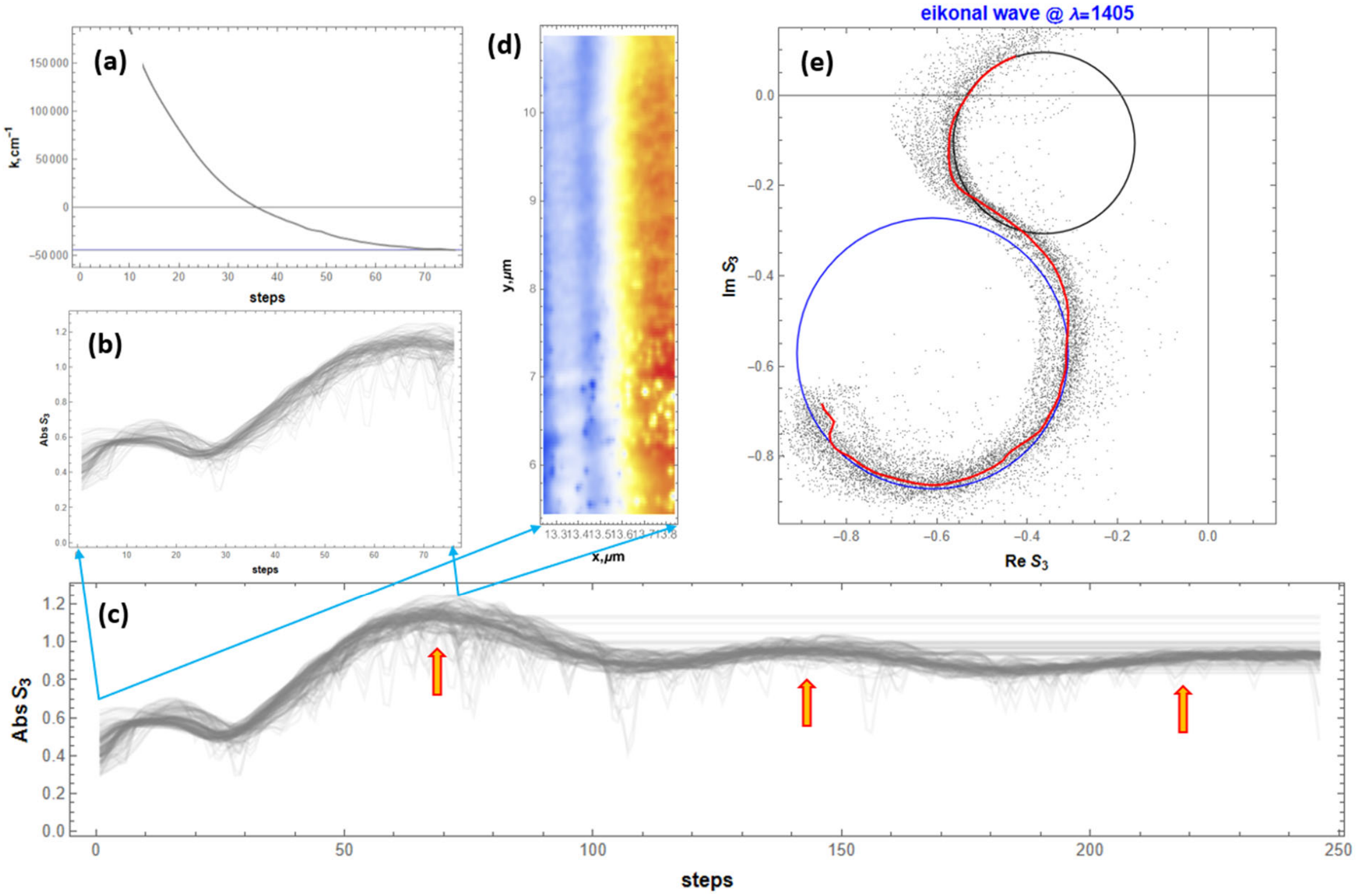

**Figure S12.** Eikonal wave analysis of a single polaritonic sSNOM map. (a) Convergence of the polaritonic phase derivative ($-i\partial\phi/\partial x$), equal to the wavevector of polariton to the right of the edge, to the "ordinary" value (blue gridline) obtained from the manual wavelength detection (cf. the distance between the arrows in panel (c)). Partial (b) and full (c) polaritonic wave profile to the right of the edge. (d) sSNOM map showing actual data, used to plot profiles in (b). (e) An Argand plot of the whole map in (d). Red curve is an average phase trajectory of a polariton wave, used to get the wavevector in (a); black and blue fitting circles allow to find and subtract the value of sSNOM background, different on two sides of the edge.

The Argand plot in Figure S12e shows a limited part of the wave pattern (note that neither blue nor black arc has more than 180° arc length). Actual area taken for analysis (a fraction of the whole map in Figure S11d) is shown in Figure S12d. Corresponding wave profiles are plotted in Figure S12b. Note that after data alignment, all profile curves collapse. For the purpose of comparison, we present also much longer profile curves in the panel (c) – that amount of data would be needed to manually determine the "ordinary" wavelength, using the distance between the maxima indicated by arrows. As one can judge the manual wavelength requires averaging over the large distance and is not free from ambiguities: below we will use manual wavevectors for comparison with the eikonal results (more than one fit is possible, thus for some maps, we will present several values of a manual wavevector).

For theoretical analysis of the polaritons in the sandwich material we used a classical model of light reflecion from a stratified medium. It is well known that the polaritonic modes of such a medium correspond to the poles of the total reflection coefficient, R, and can be visualized when plotting R vs. the frequency and wavevector of the polariton. Each layer is characterized by a frequency dependent dielectric tensor[i] (for uniaxial hBN), or a single dielectric function[ii] (for isotropic amorphous $SiO_2$), or a single value of the dielectric constant[iii] eps=11.7 (for Si, which is known to have negligible dispersion in the mid-IR region), or a two-dimensional optical conductivity[iv] (for room-temperature Dirac model for graphene, with the dependence on the doping level). Although the parameterization of the model does not change the result qualitatively, it may make the dispersion curves to slightly deviate from measured values.

Figure S13 presents spectral data used in this section: panel (a) shows Re and (negligible) Im part of the dielectric function used for amorphous $SiO_2$ layer; panel (b) shows two-dimensional optical conductivity of a monolayer graphene for $n_G=2x10^{12}$ $cm^{-2}$ doping level (best fit); since the 2d-conductivity has the units of velocity it was divided by the speed of light for making a unitless quantity; panel (c) shows Re and Im parts of both components of dielectric tensor of hBN (ordinary/in-plane, and extraordinary/out-of-plane).

We construct the propagators for electromagnetic waves in each layer as:

$$k_n = \sqrt{\varepsilon_n\, \beta^2 - q^2}$$

for the dielectric functions of isotropic media: n = 2 [Si], or n = 1 [SiO2], or n = 0 [air ($\varepsilon = 1$)], where $\beta = \frac{\omega}{c}$, here c is the speed of light, ω is the frequency, q is the polariton wavevector (in plane).

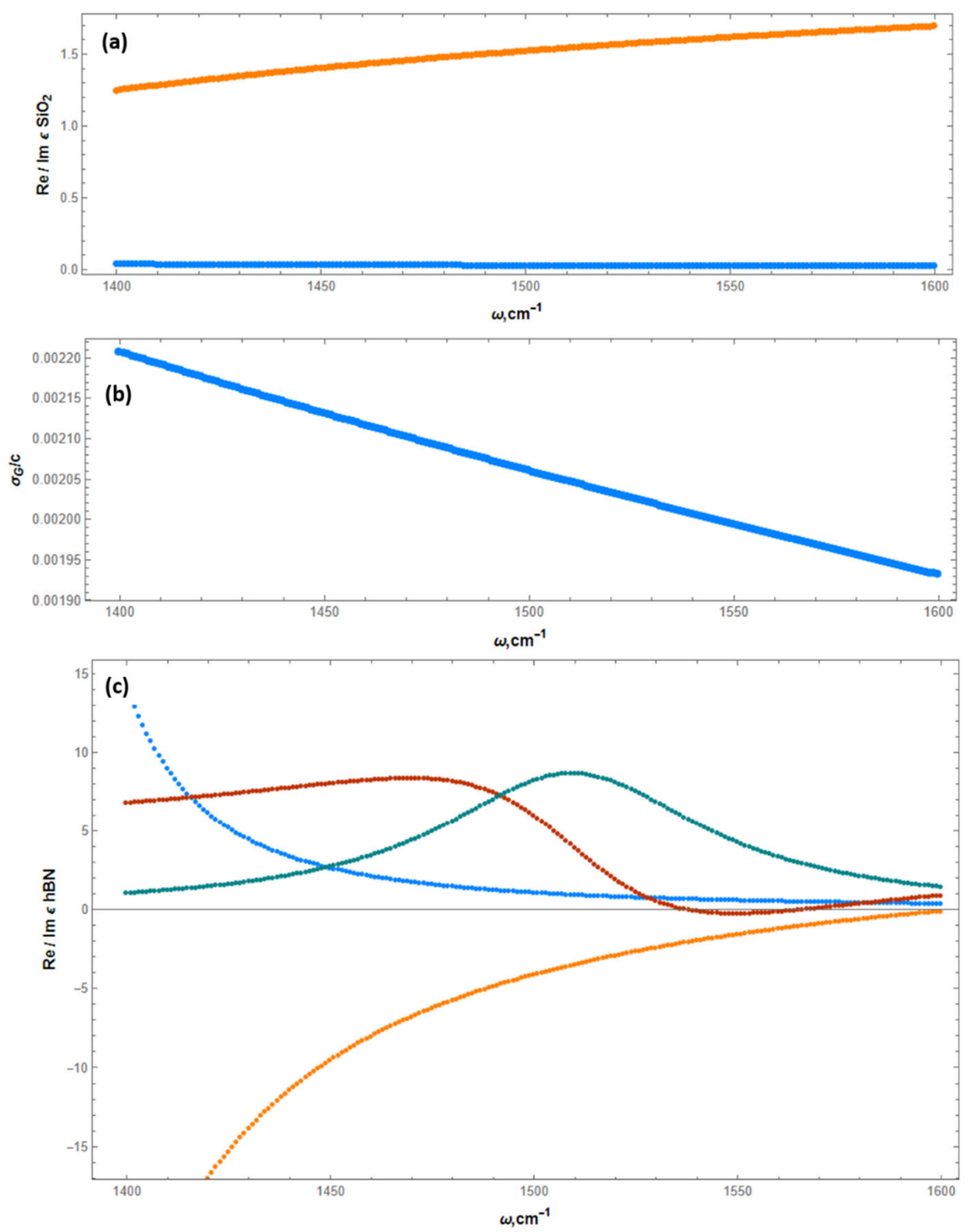

**Figure S13.** The dielectric function parameters used in this section: (a) Re (orange) and Im (blue) parts of the dielectric function for amorphous $SiO_2$ [2]; (b) two-dimensional optical conductivity of monolayer graphene [4], divided by the speed of light, for doping level: $n_G=2x10^{12}$ $cm^{-2}$; (c) Re (orange) and Im (blue) parts of ordinary/in-plane components of the dielectric tensor of hBN; Re (maroon) and Im (emerald) parts of extraordinary/out-of-plane components [1]. All functions are shown in the spectral range 1400-1600 $cm^{-1}$.

The propagator for an isotropic medium (hBN) reads as:

$$k_{hBN} = \sqrt{\varepsilon_{hBN}^{e}\,(\beta^2 - q^2/\varepsilon_{hBN}^{o})}$$

where the in-plane and out-of-plane components of the dielectric tensor of n = hBN are given by $\varepsilon_{hBN}^{o}$, and $\varepsilon_{hBN}^{e}$. Finally, we introduce the parameter for MLG with the units of propagator:

$$Q = \frac{i\,\omega}{2\pi\sigma}$$

where $\sigma$ is the optical conductivity:

$$\sigma = i\,\frac{e^2\,v_F}{\pi\,\hbar\,\omega}\sqrt{\pi\,n_G}$$

where $v_F$ ~$10^6$ cm/s is the Fermi velocity;

With these parameters, we define the reflection coefficients between layers n and m as:

$$r_m^n = \frac{k_n/\varepsilon_n - k_m/\varepsilon_m}{k_n/\varepsilon_n + k_m/\varepsilon_m}$$

where for the anisotropic medium one substitutes $\varepsilon_{hBN}^{e}$ for the dielectric function. While for graphene monolayer we need:

$$r_G = \frac{q}{q - Q}$$

The Fresnel reflection coefficient of the whole sandwich slab t-hBN/MLG/b-hBN/$SiO_2$/Si is analytically calculated using the transmission matrix approach. Each layer (except for MLG) contributes a factor T`.G.T, where G is the free propagation matrix, and T-matrices (interface matrices) are written via the $r_m^n$ coeffcicients. The total transmission matrix reads as:

$$T = \begin{bmatrix} \frac{1}{1+r_3^2} & \frac{r_3^2}{1+r_3^2} \\ \frac{r_3^2}{1+r_3^2} & \frac{1}{1+r_3^2} \end{bmatrix} \cdot \begin{bmatrix} e^{i\Delta} & 0 \\ 0 & e^{-i\Delta} \end{bmatrix} \cdot \begin{bmatrix} \frac{1}{1+r_2^1} & \frac{r_2^1}{1+r_2^1} \\ \frac{r_2^1}{1+r_2^1} & \frac{1}{1+r_2^1} \end{bmatrix} \cdot$$

$$\begin{bmatrix} e^{i\Delta_b} & 0 \\ 0 & e^{-i\Delta_b} \end{bmatrix} \cdot \begin{bmatrix} \frac{1}{1-r_1^0} & \frac{-r_1^0}{1-r_1^0} \\ \frac{-r_1^0}{1-r_1^0} & \frac{1}{1-r_1^0} \end{bmatrix} \cdot \begin{bmatrix} \frac{1-2r_G}{1-r_G} & \frac{r_G}{1-r_G} \\ \frac{-r_G}{1-r_G} & \frac{1}{1-r_G} \end{bmatrix} \cdot \begin{bmatrix} \frac{1}{1+r_1^0} & \frac{r_1^0}{1+r_1^0} \\ \frac{r_1^0}{1+r_1^0} & \frac{1}{1+r_1^0} \end{bmatrix} \cdot \begin{bmatrix} e^{i\Delta_t} & 0 \\ 0 & e^{-i\Delta_t} \end{bmatrix} \cdot \begin{bmatrix} \frac{1}{1-r_1^0} & \frac{-r_1^0}{1-r_1^0} \\ \frac{-r_1^0}{1-r_1^0} & \frac{1}{1-r_1^0} \end{bmatrix}$$

There are 4 fitting parameters we chose to be close to experimentally measured physical ones: thickness of t-hBN and b-hBN layers, taken as 5 nm and 340 nm; thickness of $SiO_2$ layer of 285 nm (nominal); doping level of MLG: $n_G = 2\times10^{12}$ $cm^{-2}$. As mentioned earlier, the results are not too sensitive to the choice of these parameters. The Fresnel coefficient is deduced from matrix elements of the total transmission matrix as:

$$R = \frac{-T_{12}}{T_{22}}$$

The density map of the imaginary part of Fresnel coefficient R shows all polaritons modes in the slab. While low-q (long-wavelength) modes have complicated structure (not to be analyzed here), the modes in the region of interest (20,000-120,000 $cm^{-1}$) are coming mostly from the waves bound to the hBN due to the spectral region of interest (1400-1600 $cm^{-1}$) is within the Reststrahlen band of hBN (see Figure S13c). One of these modes is clearly visible as a C-shape curve in Figure 14. All red symbol data points, derived from the raw sSNOM maps using the eikonal model, above ~1420 $cm^{-1}$ fall on this curve (a few points at the bottom of dispersion curve show coupling of this mode to a longer wavelength mode below it). Notably, the manual fit of the polariton wavelength (points of other color without error bars) show similar trends, though have a significant scatter and contain several outliers.

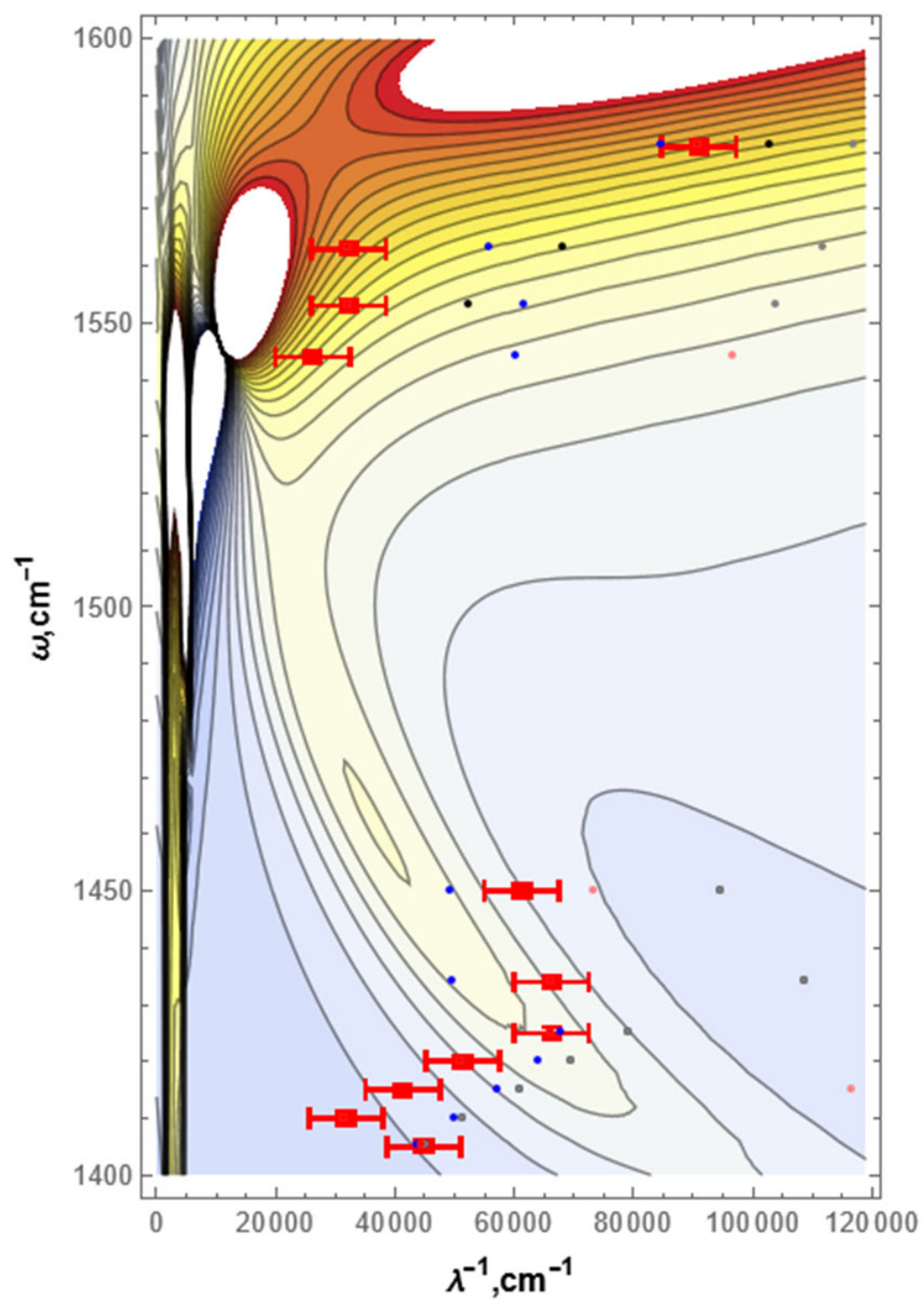

**Figure S14.** The comparison of eikonal model for sSNOM-based polariton dispersion curve (red symbols with error bars), with the manual wavelength fitting (all other points), and the theoretical prediction (density plot) Im(R) for the t-hBN/MLG/b-hBN/$SiO_2$/Si slab with the parameters described in the text.

In conclusion, here we demostrate that the eikonal method allows to determine polariton wavevectors that have quantitive agreement with the classical theory for multilayer slab polaritons.

**S9. Polariton dispersion in non-structured SiC/2D-Ag/EG films**

Figure S15 and S17 show the sSNOM maps of non-structured SiC/2D-Ag/EG films, chosen from the hyperspectral series SH1 (76 single laser frequency maps) and SH2 (62 single laser frequency maps). We stress that since the material morphology is highly non-uniform – it demonstrates multiple terrace steps, other layer non-uniformities, line and point defects – the propagation of free polariton waves is highly impeded by these morphological features that appeared at the distances much shorter than the polariton wavelength. As a result, the polaritons are frequently scattered and show a complex diffraction pattern. We provide a few typical maps in Figure S15 and S18. As one can see, no clear wave pattern can be traced in these maps, completely prohibiting simple real-space wavelength analysis (counting wavelength distance from the maps).

Nevertheless, using Fourier transform of the maps (with Gaussian filtering of the data to remove high-frequency components which could produce a significant noise) one can trace several spots in the 2D Fourier space that correspond to the reflexes of the scattered polariton waves. For complex film morphology, multiple scattering events are frequent and, therefore, in the Fourier space one can detect typically multiple harmonics of the wavevectors, that show as a series of peaks at commensurate values along the same direction in the reciprocal space.

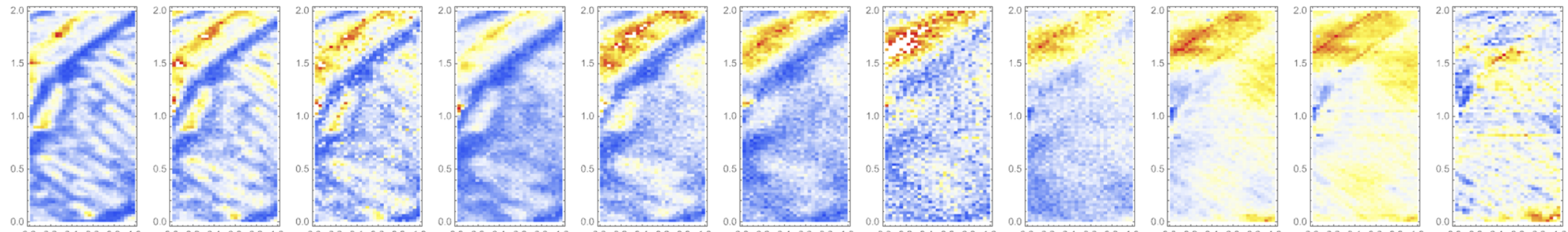

**Figure S15.** Typical raw sSNOM O3A maps of the hyperspectral series for sample SH1. Each map is taken over the same area 1um x 2um and contains 120,000 pixels (this figure does not show full resolution of the map). Total series covers the spectral range 925-1000 $cm^{-1}$, with the step of 1 $cm^{-1}$ (only few maps are shown here).

Figure S16 shows a typical 2D Fourier map of one of the images in Figure S15 – a large central peak of zero frequency reflects long-wavelength fluctuations over the map area, seen in real-space maps. The red line shows direction in the reciprocal space which we chose to plot the polariton dispersion. Multiple peaks along this line correspond to several harmonics.

Figure S17 (left) shows a hyperspectral representation of the Fourier transform of the raw sSNOM map of O3A channel, which is a single wavevector/single laser frequency component for the near-field amplitude (To process the 2D Fourier data a Gaussian filter has been applied to remove high-frequency noise; the data was then interpolated onto a uniform grid and re-sampled along a specific direction in the reciprocal space; then, the FT data for each laser wavelength of the measured hyperspectral series was combined into the density map).

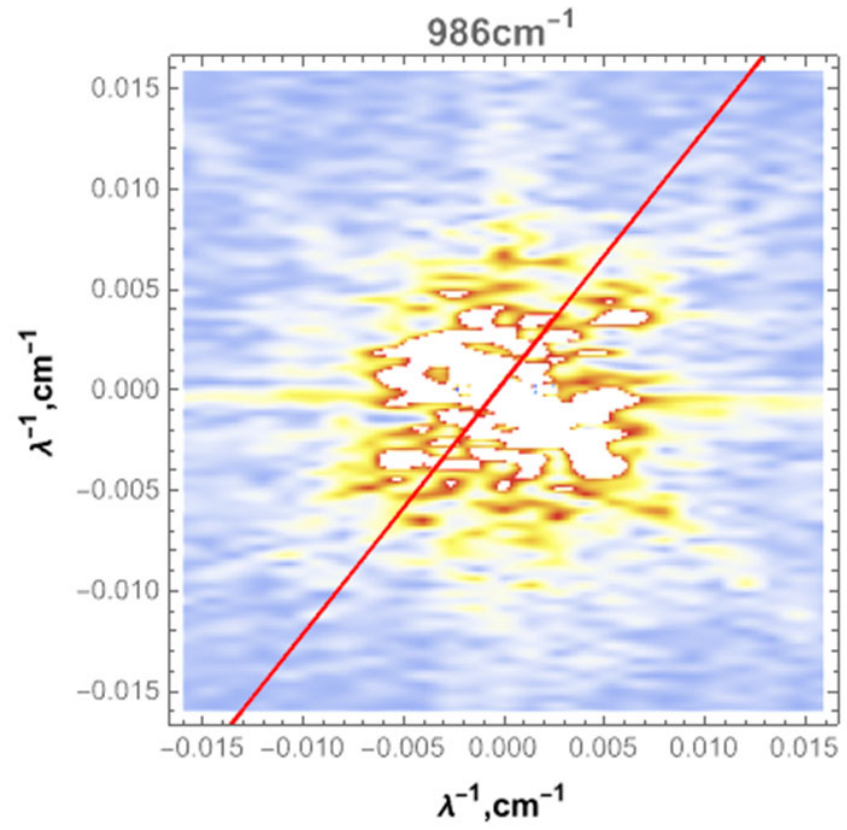

**Figure S16.** 2D Fourier map of the sSNOM O3A map at 951 $cm^{-1}$, shown in Fig. S15 (#9). Red line shows the direction for taking spectral dispersion.

One can clearly trace several features of the dispersion map: major polariton dispersion branch with a steep slope (for inverse wavelength values between 10,000 and 20,000 $cm^{-1}$); discontinuity of the dispersion of this branch near 945 $cm^{-1}$ frequency; multiple harmonics of the polaritons (due to multiple reflections) that show up as nearly parallel lines/commensurate wavevectors; notably, the amplitude of the higher harmonics decays below certain frequency.

Interestingly, Figure S17 (right) shows the hyperspectral series SH1 compared with the dispersion curve of a pQD sample from Figure 4d (main text). It can be traced that the upper branch of the pQD polariton (red dots) dispersion follows the line of "bulk" polariton – unfortunately, in this SH1 series, the spectral data for bulk sample cannot be explicitly traced in the region of interest. This problem has been resolved in 2nd hyperspectral series below. The lower branch of pQD polariton, in its upper part above 1000 $cm^{-1}$ frequency, also follows the high-k harmonic of bulk polariton (around 45,000 $cm^{-1}$).

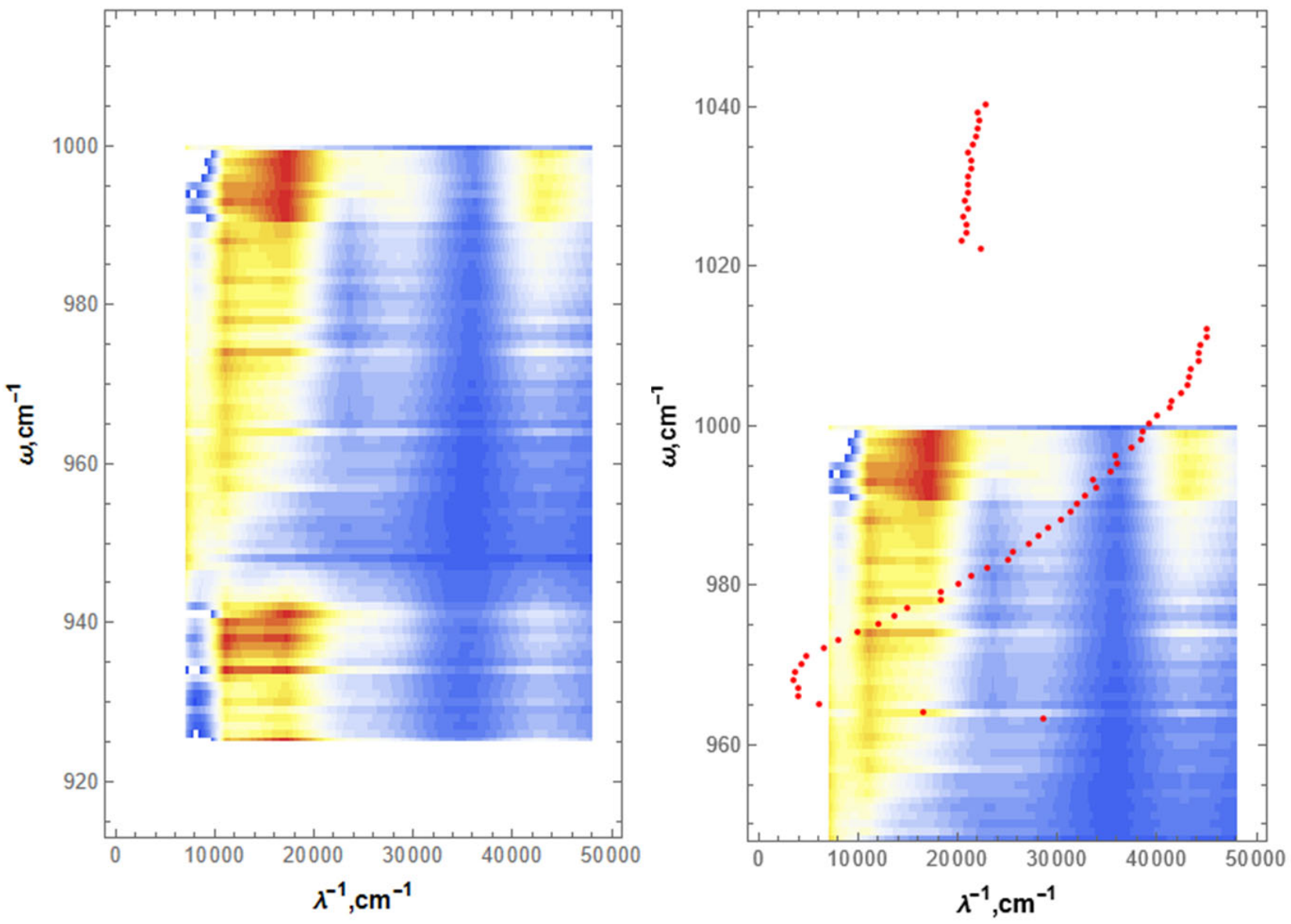

**Figure S17.** Hyperspectral sSNOM dispersion map for non-structured SiC/2D-Ag/EG film (no plasmonic dots fabricated) (hyperspectral series for sample SH1). (left) Actual measured data; (right) same map, rescaled to be overlaid with the dispersion curve from Fig. 4d (in main text).

There is a visible feature near 965 $cm^{-1}$ in pQD dispersion (a broad S-shape/inverse dispersion part of the pQD-SPP line) which is in accordance with the narrow, though resolvable feature of the bulk-SPP. Not all the features of the plasmonic confined sample and the bulk one are the same: for example, a similar narrow feature of the bulk-SPP near 974 $cm^{-1}$ has no corresponding signature for pQD.

Finally, the discontinuity feature of pQD polariton: the spectral gap between upper branch and lower branch has a correspondence with the amplitude of the high-k harmonics of the bulk-SPP: the high wavevector harmonics have negligible amplitude in the region below the pQD lower branch curve. All these observations are more clearly evidenced in next example SH2.

Figure S18 shows a few near-field maps from SH2 series, taken in the laser frequency range 916-1039 and 1084-1105 $cm^{-1}$, with the step of 3 $cm^{-1}$ (the gap is due to capability of the MIR laser we used). This is a similar non-structured SiC/2D-Ag/EG film material, though we chose a different sample to demonstrate reproducibility of the results.

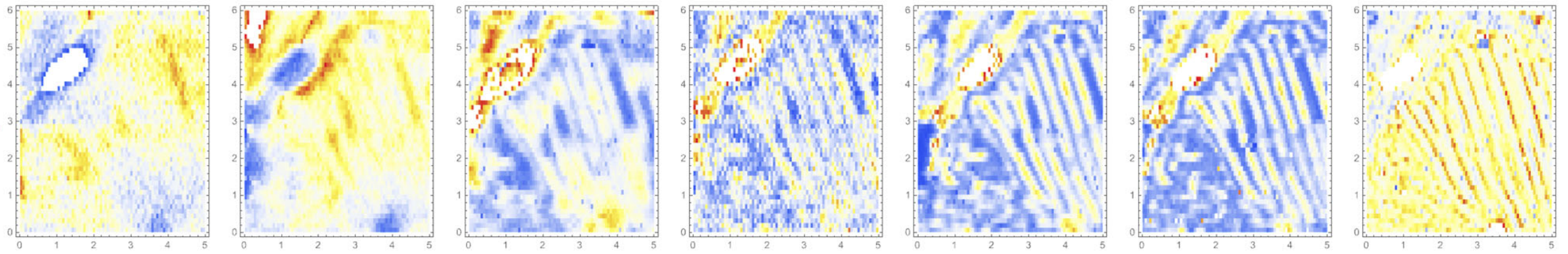

**Figure S18.** Typical raw sSNOM O3A maps of the hyperspectral series for sample SH1. Each map is taken over the same area 5um x 6um and contains 192,000 pixels (this figure does not show full resolution of the map). Total series covers the spectral range 916-1105 $cm^{-1}$, with the step of 3 $cm^{-1}$ (only few maps are shown here).

In this SH2 series one can clearly see that the pQD upper branch follows one of the bulk SPP high-k harmonic dispersion. Indeed, the pQD lower branch curve closely follows the boundary of the (upper) bulk SPP region where the amplitudes of the high-k harmonics are non-vanishing. The bulk SPP dispersion clearly demonstrates spectral features near 1025 cm$^{-1}$, which coincides with the beginning of the upper branch of the pQD polariton, and another one near 965 cm$^{-1}$, where the S-shape of pQD polariton has the minimum wavevector.

Overall, we speculate that close resemblance of the bulk SPP data in the whole studied spectral region of pQD confined polariton dispersion is due to the same physics of MIR polaritonic response of the SiC/2D-Ag/EG material, properly captured by new spectroscopic technique introduced in this paper.

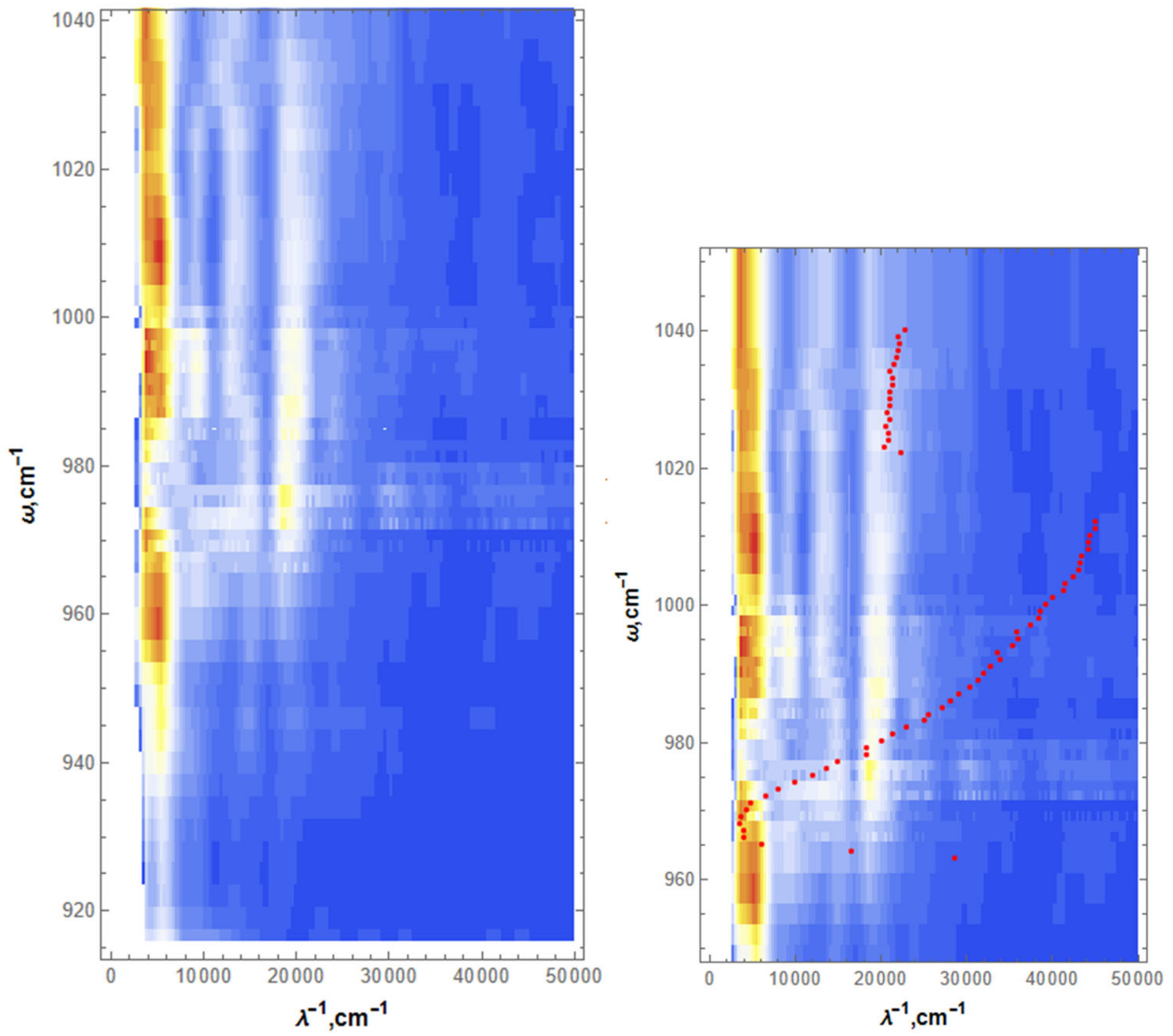

**Figure S19.** Hyperspectral sSNOM dispersion map for non-structured SiC/2D-Ag/EG film (no plasmonic dots fabricated) (hyperspectral series for sample SH2). (left) Actual measured data; (right) same map, rescaled in the same spectral region as in Fig. S17 to be overlaid with the dispersion curve from Fig. 4d (in main text).

## References

[i] Geick R, Perry CH, Rupprecht G. Normal Modes in Hexagonal Boron Nitride. Physical Review. 1966;146(2):543-7. doi: 10.1103/PhysRev.146.543.

[ii] Gervais F, Piriou B. Temperature dependence of transverse and longitudinal optic modes in the α and β-phases of quartz. Physical Review B. 1975;11(10):3944-50. doi: 10.1103/PhysRevB.11.3944.

[iii] Optical Constants of Crystalline and Amorphous Semiconductors. Sadao Adachi. Springer Science+ Business Media, LLC. Originally published by Kluwer Academic Publishers in 1999.

[iv] Woessner, A., Lundeberg, M., Gao, Y. et al. Highly confined low-loss plasmons in graphene–boron nitride heterostructures. Nature Mater 14, 421–425 (2015). https://doi.org/10.1038/nmat4169